\documentclass[preprint,pre,aps]{revtex4}
\usepackage[dvips]{graphicx}
\begin{document}
\title{Field theory for size- and charge asymmetric
 primitive model of ionic systems. Mean-field stability analysis and
 pretransitional effects}
 \author { A.Ciach$^{1}$, W.T. G\'o\'zd\'z$^{1}$  and G.Stell$^2$}
\affiliation{$^{1}$Institute of Physical Chemistry, Polish Academy of
 Sciences 01-224 Warszawa, Poland \break $^2$Department of Chemistry,
 State University of New York Stony Brook, NY 11 794-3400, USA}
\date{\today}
\begin{abstract}
The primitive model of ionic systems is investigated within a
field-theoretic description for the whole range of size- , $\lambda$,
and charge, $Z$, ratios of the two ionic species. Two order parameters
(OP) are identified, and their relations to physically relevant
quantities are described for various values of $\lambda$ and $Z$.
Instabilities of the disordered phase associated with the two OP's are
determined in the mean-field approximation (MF).  In MF a gas-liquid
separation occurs for any $Z$ and $\lambda\ne 1$. In addition, an
instability with respect to various types of periodic ordering of the
two kinds of ions is found. Depending on $\lambda$ and $Z$, one or the
other transition is metastable in different thermdynamic states.  The
instabilities found in MF represent weak ordering of the majority of
the instantaneous states, and are identified with structural loci
associated with pretransitional effects.
\end{abstract}
\maketitle 
\section{Introduction}
 Universal features of collective phenomena can be determined within
 the framework of statistical mechanics with the interaction
 potentials approximated by highly simplified generic models.  The
 generic model that allows for a prediction of general features of the
 collective phenomena in ionic systems as diverse as molten salts,
 electrolytes, room-temperature ionic liquids, as well as systems
 containing charged nanoparticles (including proteins), and/or charged
 colloidal particles, is the primitive model (PM). In the PM ions are
 represented by charged hard spheres, and the solvent (if present) is
 taken into account only through the dielectric
 constant~\cite{stell:95:0,fisher:94:0,ciach:05:0}. In real systems
 the molecular structure of the solvent and multipole moments and
 polarizability of ions and solvent molecules affect the collective
 phenomena in the way that depends significantly on the specific
 properties of the ions and the solvent~\cite{zemb:04:0,tavares:04:0},
 and is far from being understood. To identify and analyze specific
 effects in particular cases, however, one should know, in the first
 place, the phase diagram and structure in the generic case, where
 such effects are not present.

  Despite the simplicity of the interaction potential in the PM, the
  dependence of the phase diagrams and the correlation functions on
  the charge,
\begin{equation}
\label{cZ}
Z=\frac{e_+}{|e_-|},
\end{equation}
and the diameter,
\begin{equation}
\label{c}
 \lambda=\frac{\sigma_+}{\sigma_-},
\end{equation}
ratios of the two ionic species, respectively, is a largely unexplored
problem.  Until very recently the theoretical studies concentrated
mainly on the restricted primitive model
(RPM)\cite{stell:95:0,fisher:94:0,levin:02:0,ciach:05:0}, where
$Z=\lambda=1$. In the RPM a gas-liquid separation occurs at low
densities, and the bcc ionic crystal occurs at higher densities and
sufficiently low temperatures. At still higher densities the fcc
crystal with and without substitutional order at low and at high
temperatures respectively is stable~\cite{ciach:05:0,vega:03:0}.

 Although much studied, the RPM is a quite artificial model, in the
 sense that no real ionic fluid has anions and cations of {\it
 exactly} the same size, although the size disparity is rather small
 in some, such as KCL.  When one assumes full anion-cation symmetry,
 as in the RPM, there is a remarkable decoupling of the
 Ornstein-Zernike (OZ) integral equations that describe the
 density-density correlation function and the charge-charge
 correlation function, respectively.  The two correlation functions
 are still indirectly coupled through the "closure relations" that
 must be added to the OZ equations to yield a closed set of equations,
 and the different approximations that have come into standard use in
 solving the primitive-model OZ equations-- the mean spherical
 approximation, the hypernetted chain approximation, etc.-- are
 defined by different closure relations~\cite{stell:95:0}. The total
 and the direct correlation function in the liquid theory are directly
 related to the (connected) correlation function and the vertex
 function respectively in the field theory (FT), and similar
 decoupling of the analogs of the OZ equations in the FT approach is
 found~\cite{ciach:05:0}.

As soon as there is any degree of asymmetry, however, the
density-density and charge-charge correlation functions are directly
coupled, as are the OZ integral equations for these two
functions~\cite{stell:95:0}, as well as their analogs in the 
FT~\cite{ciach:05:0,ciach:06:0}.  We shall not go through the details
of this case here, and the interested reader will find those details
in Section 2.2 of Ref.\cite{stell:95:0}. There is one aspect of the
results there that is worth noting here explicitly, however.  If there
is only a very small degree of asymmetry-- for example, if the
effective diameters of the anions and cations are very nearly the same
in a 1-1 electrolyte, one would expect that the RPM equations would be
"almost" satisfied in some sense, despite the direct coupling one
finds in the asymmetric case. This turns out to be the case-- one
finds that deviations from the RPM result only appear very close to
the critical point for very small asymmetries. The smaller the
deviations, the closer one must be to the critical point (cp) to see
them.  Recent
theoretical~\cite{kalyuzhnyi:00:0,zuckerman:01:0,artyomov:03:0,aqua:04:0,raineri:00:0,qin:04:0,aqua:05:0,zhou:05:0,patsahan_mryglod:06}
and simulation~
\cite{romero:00:0,yan:01:0,yan:02:0,panag:02:0,cheong:03:0,rescic:01:0,linse:01:0} results show that the PM with a small and moderate asymmetry 
exhibits qualitatively the same critical behavior as the RPM. In the
asymmetric case the cp is shifted to lower temperatures and number
densities (but higher volume
fractions)~\cite{romero:00:0,yan:01:0,yan:02:0,panag:02:0,cheong:03:0,rescic:01:0,linse:01:0,aqua:05:0,zhou:05:0,patsahan_mryglod:06}. We
expect to make further contact between our FT treatment for asymmetric
ionic fluids and the general results of Ref.\cite{stell:95:0}
concerning the critical behavior in future work.  As far as we know,
the effect of the size- and charge asymmetry on the gas-solid and
liquid-solid transitions attracted much less attention so far.

Despite our general insight into the asymmetric case that has come out
of the work of Ref.\cite{stell:95:0}, and intensive studies of highly
charged colloids
~\cite{derjaguin:41:0,verwey:43:0,warren:00:0,roij:99:0,ise:83:0,
ise:99:0,ito:94:0,belloni:97:0,belloni:00:0,tata:97:0,kjellander:97:0},
the case of the extreme asymmetry, $Z,\lambda\to\infty$, is not fully
understood.  This is because on the theoretical side the effect of
approximations and assumptions on the results may be significant.  In
experiments~\cite{ise:83:0,ise:99:0,arora:98:0,ito:94:0,tata:97:0} the
interactions differ from the PM, and the role of specific interactions
and/or of the solvent properties is not known in full
detail. Moreover, results may be sensitive to impurities,
nonequilibrium effects etc..  Last but not least, it is very difficult
to obtain reliable simulation results for the phase diagram when
$Z,\lambda\to\infty$. To ensure charge neutrality $Z$ anions must be
present per each cation, and in studies of collective phenomena the
number of cations $N_+$ must be large. Typical experiments correspond
to $Z\sim 10^4$, i.e. to $\sim 10^4N_+$ ions in the simulation box.
Because of the above reasons, there is no consensus concerning the
phase behavior in the PM in the case of the extreme asymmetry, and the
results of some experiments~\cite{ise:99:0,tata:97:0} are considered
as controversial~\cite{kjellander:97:0,belloni:00:0}. The controversy
concerns mainly the occurrence of the gas-liquid separation at room
temperature in the case of monovalent counterions. Simulations
indicate that at room temperatures the gas-liquid separation for
monovalent counterions does not occur, but occurs for divalent
counterions~\cite{belloni:00:0}. Preliminary field-theoretic
results~\cite{ciach:06:0} indicate that when the size and the charge
of the cation are several orders of magnitude larger than the size and
the charge of the anion, the gas-liquid type phase separation is
preempted by a transition between gas and a colloidal crystal for a
large range of temperatures, including  room
temperature~\cite{ciach:06:0}. The predictions of
Ref.~\cite{ciach:06:0} are limited to the mean field (MF)
approximation, and the effect of fluctuations on the transition lines
remains to be determined along the same lines as in the case of the
RPM~\cite{ciach:06:2}. The phase behavior for higher volume fractions
is also not quite clear. Many studies suggest a transition between the
bcc and fcc crystals, and next a gel or glass formation when the
volume fraction of colloidal particles
increases~\cite{choudhury:95:0,lowen:94:0,arora:98:0}. Additional
counterions and coions significantly influence the phase behavior~
\cite{warren:00:0,yu:04:0,roij:99:0,levin:02:0,lowen:94:0,arora:98:0}. Note
that the above phase behavior is quite different than the behavior in
the RPM.

 The way the global phase diagram evolves when the asymmetry
 parameters are varied has not been systematically investigated yet.
 The values of $Z$ and $\lambda$ have a strong effect on formation of
 ordered periodic structures of a crystal type. Coulomb interactions
 support structures in which the positive charges are compensated by
 the negative charges in regions as small as possible. On the other
 hand, precise compensation of charges in small regions may lead to an
 increase of entropy.  The electrostatic and the entropic
 contributions to the grand potential both depend on the geometrical
 constraints. These constraints are due to a packing of $Z$ spheres of
 a diameter $\sigma_-$ and one sphere of a diameter
 $\sigma_+=\lambda\sigma_-$ into a charge-neutral aggregate, and a
 possibility of periodically repeating this aggregate in space. Such
 an ordered structure is favored compared to the disordered phase when
 the entropic contribution is sufficiently small compared to the
 electrostatic energy. Combination of the geometric constrains and the
 charge-neutrality condition for some values of $Z$ and $\lambda$ may
 promote ordering, while in other cases may suppress ordering. The
 competition between the periodic ordering and the phase separation
 also depends on $Z$ and $\lambda$. Thus, rich and complex phase
 behavior in the parameter space $(\zeta,T^*;Z,\lambda)$ may be
 expected, where $\zeta$ and $T^*$ are the volume fraction of ions and
 temperature in the standard reduced (dimensionless)
 units~\cite{stell:95:0,ciach:05:0}, defined in the next section. Depending on the values of $Z$ and $\lambda$, phase
 equilibria on the $(\zeta,T^*)$ phase diagram may include some (or
 all) of the following transitions: gas-crystal, gas-liquid,
 liquid-crystal, crystal-glass, crystal-gel or transition between
 different crystalline phases. For different $Z$ and $\lambda$ the
 above phase equilibria may occur in quite different parts of the
 $(\zeta,T^*)$ phase diagram.

Some insight into the dependence of the phase behavior on $Z$ and
$\lambda$ can be gained from studies of pretransitional effects in the
disordered phase.  Such effects may be divided into two categories:
effects associated with the gas-liquid separation and effects
associated with the crystallization.  Theoretical
studies~\cite{stell:95:0,ciach:05:0,ciach:06:1,aqua:05:0,zhou:05:0}
emphasize the role of cluster formation for the gas-liquid separation
in ionic systems.  Recent Monte Carlo
(MC)~\cite{hribar:97:0,hribar:00:0,linse:01:0,rescic:01:0,cheong:03:0,yan:02:0,yan:02:1,yan:01:0,weis:98:0,panag:02:0},
molecular dynamics (MD)~\cite{spohr:02:0} and Brownian dynamics
(BD)~\cite{jardat:01:0} simulation studies provide interesting
information on clustering in the size- and charge asymmetric
PM.  In
Refs.\cite{hribar:97:0,hribar:00:0,spohr:02:0,jardat:01:0} the
simulations were performed at room temperature for several values of
the volume fraction of the larger ions, for several sizes of both
ionic species (size ratio ~5-10), and for several values of the charge
on the larger ion (~10-20 elementary charges).  The cluster formation
was studied for mono- di- and trivalent counterions for fixed values
of the remaining parameters.  In equilibrium the instantaneous
structures are much more uniform in the case of monovalent counterions
than in the other cases, where 'living' clusters containing two or
more large ions are formed ~\cite{hribar:00:0,jardat:01:0}. For
trivalent counterions a substantial fraction of the clusters was found to be
neutral.  Attractions between the neutral clusters are similar to the
attractions between molecules, and may lead to the phase
separation. The dependence on valency of counterions for a fixed
charge on the larger ion and for fixed temperature is equivalent to a
dependence on $Z$ and $T^*$. The question of the region in the
parameter space $(\zeta,T^*;Z,\lambda)$ corresponding to a formation
of neutral clusters remains open.

  Pretransitional effects associated
with weak long-range ordering are a subject of the present study. By
analogy with the {\it short-range ordering of the instantaneous
states} described above, we expect that such effects include {\it
long-range ordering of the instantaneous states}. In the long-range
ordered instantaneous states a local deviation from the uniform
distribution of ions is periodically repeated in space.

In this work we present an overview of the collective phenomena in the
PM for the whole range of $Z$ and $\lambda$. Our FT analysis allows us to
identify the dominant deviations from the uniform distribution of
ions.  We find structural lines that separate the stability region of
the disordered phase on the $(\zeta,T^*)$ phase diagram in the
high-temperature part where the homogeneous instantaneous states are
more probable than periodic states with small amplitude, and in the
low-$T^*$ part, where the majority of the instantaneous states exhibit
weak long-range order. At the structural line the amplitude of the
periodic deviation from the uniform distribution of ions in the most
probable instantaneous states increases continuously from zero. We
stress that the structural line is neither a phase transition in the
thermodynamic sense, nor a true spinodal.  Determination of the
probability of the periodic instantaneous states with large
amplitudes, as well as determination of the phase transitions requires
further studies, for which our analysis should serve as a starting
point.

In sec.II we describe the field theory for the PM. We introduce the
concept of the structural line and show that this line coincides with
the MF approximation for the spinodal line.  In sec.III general
expressions for the eigenmodes and boundary of stability of the
uniform phase in MF (structural lines) are given. The nature of the
eigenmodes is discussed, and the dominant instantaneous states are
identified in the same section. Sec.IV is devoted to the case of equal
sizes. MF-spinodal (structural) lines associated with the gas-liquid
separation are described in detail in sec.V, and in sec.VI the
structural lines associated with periodic ordering are
discussed. Explicit results for representative values of $Z$ and
$\lambda$ are also presented in these two sections. A short summary is
given in sec.VII.
\section{Field-theoretic description of the primitive model}
\subsection{Coarse graining of the PM}
We consider a mixture of positively charged ions of a charge $e_+$ and
diameter $\sigma_+$, and negatively charged ions of a charge
$e_-=-|e_-|$ and diameter $\sigma_-$ in a structureless,
incompressible solvent. In the PM the interaction
potential of a pair $\alpha,\beta$, where the Greek indices denote $+$
or $-$, is infinite for distances smaller than
$\sigma_{\alpha}+\sigma_{\beta}$, i.e. we assume hard-core repulsions.
The electrostatic potentials $V_{\alpha\beta}({\bf r}_1-{\bf r}_2)$
between different pairs of ions $\alpha,\beta $ are the following
\begin{equation}
\label{V++}
V_{\alpha\beta}(r)=
\frac{e_{\alpha}e_{\beta}}{Dr}\theta(r-\sigma_{\alpha\beta}),
\end{equation}
where 
\begin{equation}
\sigma_{\alpha\beta}=\frac{1}{2}(\sigma_{\alpha}+\sigma_{\beta})
\end{equation}
 and $D$ is the
dielectric constant of the solvent.  The $\theta$-functions above
prevent from the contributions to the electrostatic energy coming from
overlapping hard spheres, i.e.  we do not include the electrostatic
self-energy. 

In the field-theoretic approach we consider local densities of the
ionic species, $\rho_{+}({\bf r})$ and $\rho_{-}({\bf r})$, and we
include only smooth functions. For particular fields $\rho_{+}({\bf
r}),\rho_{-}({\bf r})$ we assume that the electrostatic energy assumes
the form
\begin{equation}
\label{U}
U[\rho_+,\rho_-]=\frac{1}{2}\int d{\bf r}_1\int d{\bf r}_2  
\rho_{\alpha}({\bf r}_1)V_{\alpha\beta}({\bf r}_1-{\bf r}_2)
\rho_{\beta}({\bf r}_2),
\end{equation}
where summation convention for the Greek indices is used here and below. The probability
that the local densities assume a particular form $\rho_{+}({\bf
r}),\rho_{-}({\bf r})$ is proportional to $\exp[-\beta(
U[\rho_+,\rho_-] -\mu_+N_+-\mu_-N_-)]$, where $N_+=\int d{\bf
r}\rho_{+}({\bf r}) $ and $N_-=\int d{\bf r}\rho_{-}({\bf r})$ are the
number of positive charged and negative charged ions respectively.
 The chemical potentials of the two ionic species, $\mu_+$
and $\mu_-$, are not independent, and have to be consistent with the
requirement of  charge neutrality,
\begin{equation}
\label{neut}
\int d{\bf r}\rho_+({\bf r})e_+=\int d{\bf r}\rho_-({\bf r})|e_-|.
\end{equation}
Whereas the charge neutrality condition must be obeyed in macroscopic
regions, it can be violated locally, in regions containing a small
number of ions.  The probability that the local densities assume the
form $\rho_{+}({\bf r}),\rho_{-}({\bf r})$ is also proportional to the
number of all microscopic states compatible with $\rho_{+}({\bf
r}),\rho_{-}({\bf r})$. The number of all states can be written in the
form $\exp(\beta TS)$, where $\beta=(kT)^{-1}$, and where $S$, $T$ and
$k$ are the entropy, temperature and the Boltzmann constant
respectively. From the above it follows that the Boltzmann factor assumes the form
\begin{equation}
\label{bolfac}
p[\rho_+,\rho_-]=\Xi^{-1}\exp(-\beta \Omega^{MF}[\rho_+,\rho_-]),
\end{equation}
 where 
\begin{equation}
\label{Xi}
\Xi=\int D\rho_+\int D\rho_-\exp(-\beta \Omega^{MF}[\rho_+,\rho_-]),
\end{equation}
and
\begin{equation}
\label{Omega}
\Omega^{MF}[\rho_+,\rho_-] =F_h[\rho_+,\rho_-] + U[\rho_+,\rho_-] 
-\mu_+\int d{\bf r} \rho_+({\bf r})-\mu_-\int d{\bf r} \rho_-({\bf r}) 
\end{equation}
is the grand potential in the system where the local concentrations of
the two ionic species are constrained to have the forms $\rho_{+}({\bf
r}),\rho_{-}({\bf r})$. In Eq.(\ref{Omega}) $F_h=-TS$ is the Helmholtz free energy
 of  the hard-core reference system. In the local density approximation
$F_h[\rho_+,\rho_-]=\int d{\bf r} f_h(\rho_+({\bf r}),
\rho_-({\bf r}))$, and $f_h$ consists of the ideal-gas contribution
plus the excess free-energy density of hard-spheres with different
diameters $f_h^{ex}$,
\begin{equation}
\label{fh}
\beta f_h(\rho_+({\bf r}),\rho_-({\bf r})) 
=\sum_{\alpha}\rho_{\alpha}({\bf r})
\big(\ln (\Lambda^3_{\alpha}\rho_{\alpha}({\bf r}))-1\big) + \beta
f_h^{ex}(\rho_+({\bf r}),\rho_-({\bf r})) .
\end{equation}
The $\Lambda_{\alpha}$ is the thermal de Broglie wavelength of species
$\alpha$, $\alpha =+,-$.  Here for $f_h^{ex}$ we assume the
Percus-Yevick compressibility route to the Helmholtz free-energy of a
hard-sphere mixture with size asymmetry~\cite{lebowitz:64:0}. The
above theory is equivalent to an approximate
version~\cite{patsahan:06:3} of the exact collective-variables
theory~\cite{yukhnovskii:58:0,yukhnovskii:87:0}.

 As a length unit we
choose $\sigma_{+-}$.
 The dimensionless densities and
inverse temperature are chosen as
\begin{equation}
\label{def}
\rho_+^*=\rho_+\sigma_{+-}^3,\hskip1cm\rho_-^*=\rho_-\sigma_{+-}^3,
\hskip1cm \beta^*=\beta\frac{e_+|e_-|}{D\sigma_{+-}}.
\end{equation}
 As thermodynamic variables we
choose $T^*=1/\beta^*$ and 
 the
volume fraction of all ions,
\begin{equation}
\label{zeta}
\zeta=\frac{\pi}{6}(\rho_{0+}\sigma_+^3+\rho_{0-}\sigma_-^3).
\end{equation}
The volume fraction of all ions is related to 
the quantity proportional to the number
density of all ionic species,
\begin{equation}
\label{s}
s=\frac{\pi}{6}(\rho_{0+}^*+\rho_{0-}^*)=\sigma_{+-}^3(\bar N_++\bar N_-)/V
,
\end{equation}
 where $V$ is the volume and $\bar N_{\pm}$ denote the average numbers
 of positive and negative charges for given chemical potentials. The
 relation between $\zeta$ and $s$ is given by
\begin{equation}
\label{zetas}
\zeta=s\big[1+3\delta(\delta-\nu)-\delta^3\nu\big].
\end{equation}
 We also define the parameters describing 
 the charge- and size  asymmetry,
\begin{equation}
\label{delta}
\nu=\frac{Z-1}{Z+1}
\end{equation}
and 
\begin{equation}
\label{epsilon}
\delta=\frac{\lambda-1}{\lambda+1}.
\end{equation}
 For $\delta\nu>0$
($\delta\nu<0$) the charge at the larger ion is larger (smaller).
Because of the symmetry, we limit ourselves to $\delta>0$ and consider
$-1\le \nu\le 1$.  

Let us focus on homogeneous instantaneous states corresponding to the
extremum of $\Omega^{MF}$. The densities in the  uniform states, $\rho^*_{\alpha}({\bf r})=const.$, corresponding to
$\delta\Omega^{MF}/\delta\rho^*_{\alpha}=0$, will be denoted by
$\rho_{0\alpha}^*$.  At sufficiently
high $T^*$, homogeneous instantaneous states correspond to the global
maximum of the Boltzmann factor (\ref{bolfac}), i.e. to the global
minimum of $\Omega^{MF}$. At low temperatures the extremum of
$\Omega^{MF}$ may correspond to either a local minimum, a saddle point or to a
maximum. For small deviations of $\rho_+^*({\bf x})$ and
$\rho_-^*({\bf x})$ from $\rho_{0+}^*$ and $\rho_{0-}^*$ respectively,
the functional (\ref{Omega}) can be expanded. The expansion can be
truncated for $\rho^*_{\alpha}- \rho^*_{0\alpha}\to 0$. We write
$\Omega^{MF}$ in the form
\begin{equation}
\label{OmF}
\Omega^{MF}=\Omega_0+\Omega_2+\Omega_{int},
\end{equation}
where $\Omega_0=\Omega^{MF}[\rho_{0+}^*,\rho_{0-}^*]$ is the value of
the functional at the extremum, and $\Omega_2$ denotes the Gaussian
part of the functional. In Fourier representation we have
\begin{equation}
\label{OmF2}
\beta\Omega_2[\Delta \tilde\rho_{+}({\bf k}),\Delta \tilde\rho_{-}({\bf k})]
=\frac{1}{2}\int\frac{d {\bf k}}{(2\pi)^3}
\Delta \tilde\rho_{\alpha}(-{\bf k})\tilde C^0_{\alpha\beta}({\bf k})
\Delta \tilde\rho_{\beta}({\bf k}) ,
\end{equation}
where $\Delta \tilde\rho_{\alpha}({\bf k})$ is the Fourier transform
of $\Delta \rho_{\alpha}^*({\bf
r})=\rho_{\alpha}^*({\bf r})-\rho_{0\alpha}^*$.  In Eq.(\ref{OmF2})
and hereafter the wave numbers are in $\sigma_{+-}^{-1}$ units. The
second functional derivative of $\Omega^{MF}[\rho_{+}^*,\rho_{-}^*]$
at $\rho_{\alpha}^*=\rho_{0\alpha}^*$ consists of two terms,
\begin{equation}
\label{Calphabeta}
\tilde C^0_{\alpha\beta}({\bf k})=a_{\alpha\beta} +\beta \tilde
V_{\alpha\beta}(k).
 \end{equation}
The first term is given by
\begin{equation}
\label{aab}
a_{\alpha\beta}=\frac{\partial \beta f_h}{\partial \rho_{\alpha}^*
\partial \rho_{\beta}^*},
\end{equation}
where the derivative is taken at $\rho_{\alpha}^*=\rho_{0\alpha}^*$.
Explicit expressions for $a_{\alpha\beta}$ are given in Appendix A. We
have used the chemical potentials for the asymmetric hard-sphere mixture
obtained in the Percus-Yevick approximation in
Ref.\cite{lebowitz:64:0}. The second term in $\tilde
C^0_{\alpha\beta}({\bf k})$ is the Fourier transform of the potential
(\ref{V++}) and we find
\begin{equation}
\label{Vf++}
\beta \tilde V_{++}(k)=\beta^*Z\frac{4\pi \cos(kr_+)}{k^2}
\end{equation}
\begin{equation}
\label{Vf--}
\beta \tilde V_{--}(k)=\beta^*Z^{-1}\frac{4\pi \cos(kr_-)}{k^2}
\end{equation}
\begin{equation}
\label{Vf+-}
\beta \tilde V_{+-}(k)=-\beta^*\frac{4\pi \cos k}{k^2}
\end{equation}
where
\begin{equation}
\label{rpm}
r_{\pm}=\frac{\sigma_{\pm}}{\sigma_{+-}}=1\pm \delta.
\end{equation}

The higher-order part of
the functional (\ref{OmF}) can be written in the form
\begin{equation}
\label{OmFint}
\beta\Omega_{int}=\int d{\bf r}\Bigg[
\frac{a_{\alpha\beta\gamma}}{3!}\Delta\rho_{\alpha}({\bf r})
\Delta\rho_{\beta}({\bf r})\Delta\rho_{\gamma}({\bf r})+
\frac{a_{\alpha\beta\gamma\nu}}{4!}\Delta\rho_{\alpha}({\bf r})
\Delta\rho_{\beta}({\bf r})\Delta\rho_{\gamma}({\bf r})
\Delta\rho_{\nu}({\bf r})\Bigg]+ ...,
\end{equation}
where 
\begin{equation}
\label{aabc}
a_{\alpha\beta\gamma}=
\frac{\partial a_{\alpha\beta}}{\partial\rho^*_{\gamma}}\hskip1cm{\rm and}
\hskip1cm
{a_{\alpha\beta\gamma\nu}}=
\frac{\partial a_{\alpha\beta\gamma}}{\partial\rho^*_{\nu}}
\end{equation}
and the derivatives are taken at $\rho^*_{\alpha}=\rho^*_{0\alpha}$.

The RPM limit of the above general model corresponds to
$\delta=\nu=0$ and was studied before in
Ref.\cite{ciach:00:0,ciach:05:0}. The opposite, colloid limit of
$\delta\to 1$, $\nu\to 1$ ($\lambda,Z\to\infty$) with $s=O(Z^0)$ 
(i.e. $\zeta=O(Z^{-1})$) is
described in Ref.\cite{ciach:06:0}. 

\subsection{Correlation and vertex functions and their generating functionals}
In the coarse-grained description the densities $\rho^*_{\alpha}({\bf
r})$ play the  role of the microstates. Thus, the standard definitions of thermodynamic potentials
and correlations functions can be applied. In particular, the
grand-thermodynamic potential in a presence of external fields
$J_+({\bf r})$ and $J_-({\bf r})$ is given by
\begin{equation}
\label{OOm}
\Omega[J_+({\bf r}),J_-({\bf r})]=-kT\log\Xi[J_+({\bf r}),J_-({\bf r})].
\end{equation}
where
\begin{equation}
\label{gj}
\Xi[J_+({\bf r}),J_-({\bf r})]=
\int D\rho_+\int D\rho_-\exp\Big[-\beta \Big(\Omega^{MF}[\rho_+,\rho_-]-
\int_{\bf r}J_{\alpha}({\bf r})\rho_{\alpha}({\bf r})
\Big)\Big].
\end{equation}
The external fields may play an important role in some experimental
cases. For systems where no external fields are present, as in our case, they
are introduced as auxiliary fields for computational reasons. The
Legeandre transform
\begin{equation}
\label{F}
F[\bar\rho_+({\bf r}),\bar\rho_-({\bf
r})]=\Omega[J_+({\bf r}),J_-({\bf r})]+\int_{\bf r}J_{\alpha}({\bf
r})\bar\rho_{\alpha}({\bf r}),
\end{equation}
where $\bar\rho_{\alpha}=-\delta\Omega[J_+({\bf r}),J_-({\bf
r})]/\delta J_{\alpha}$, is the density functional. The
$-\Omega[J_+({\bf r}),J_-({\bf r})]$ and $F[\bar\rho_+({\bf
r}),\bar\rho_-({\bf r})]$ are the generating functionals for the
correlation and the vertex functions
respectively~\cite{evans:79:0,zinn-justin:89:0}.  The two-point vertex
function
$C_{\alpha\beta}=\delta^2F/
\delta\bar\rho_{\alpha}\delta\bar\rho_{\beta}$
is related to the analog of the direct correlation function
$c_{\alpha\beta}$. In Fourier representation this relation has the
form~\cite{evans:79:0}
\begin{equation}
\label{direct}
\tilde C_{\alpha\beta}(k)=
\frac{\delta^{Kr}_{\alpha\beta}}{\langle\rho_{\alpha}\rangle}
-\tilde c_{\alpha\beta}(k),
\end{equation}
where $\delta^{Kr}_{\alpha\beta}$ is the Kronecker symbol, and
$\langle\rho_{\alpha}\rangle$ is the average density of the specie
$\alpha$.  In terms of the vertex functions the FT analogs of the OZ
equations take the simple form
\begin{equation}
\label{OZ}
\tilde C_{\alpha\beta}(k)\tilde G_{\beta\gamma}(k)=
\delta^{Kr}_{\alpha\gamma}, 
\end{equation}
where the Fourier transform $\tilde G_{\alpha\beta}(k)$ of the
two-point correlation function,
\begin{equation}
\label{GG}
G_{\alpha\beta}(r)=\langle\Delta
\rho_{\alpha}({\bf 0})\Delta\rho_{\beta}({\bf r})\rangle
=-\frac{\delta^2
\Omega}{\delta J_{\alpha}({\bf 0})\delta J_{\beta}({\bf r})},
\end{equation}
 is related
to the analog of the total correlation function  $h_{\alpha\beta}$ by~\cite{evans:79:0}
\begin{equation}
\label{total}
\tilde G_{\alpha\beta}(k)=
\delta^{Kr}_{\alpha\beta}\langle\rho_{\alpha}\rangle
+\tilde h_{\alpha\beta}(k)\langle\rho_{\alpha}\rangle
\langle\rho_{\beta}\rangle.
\end{equation}
  The FT approach with the local-density approximation (\ref{fh}) is
  designed for a description of the long range ordering that determines
  phase transitions. Due to the coarse-graining (or 'smearing' of the
  hard-spheres structure) the correlation functions for distances
  $< \sigma_++\sigma_-$ are not correctly reproduced.
\subsection{Mean-field approximation}
In practice the exact forms of $\Omega$ and $F$ cannot be obtained. In
order to adopt standard approximations to the PM, let us focus on the
most probable instantaneous distributions of ions
$\rho^*_{0\alpha}({\bf r})$, corresponding to the global maximum of
the Boltzmann factor (in the presence of the external fields the
latter is proportional to the integrand in Eq.(\ref{gj})). At low
$T^*$ the
most probable distributions may not be uniform; in particular,
crystalline phases are characterized by distributions of ions that are
periodic in space. We rewrite Eqs.(\ref{OOm}) and (\ref{F}) in the form
\begin{equation}
\label{gpf} 
\Omega[J_+,J_-]=\Omega^{MF}[\rho^*_{0+}({\bf r}),\rho^*_{0-}({\bf r})]
-\int_{\bf r}J_{\alpha}({\bf
r})\rho^*_{0\alpha}({\bf r})
-kT\log\Delta\Xi[J_+,J_-],
\end{equation}
and
\begin{equation}
\label{gpf1}
F[\bar\rho_+,\bar\rho_-]=
\Omega^{MF}[\rho^*_{0+}({\bf r}),\rho^*_{0-}({\bf r})]+
\int_{\bf r}J_{\alpha}({\bf
r})\big(\bar\rho_{\alpha}({\bf r})-\rho^*_{0\alpha}({\bf r})\big)
-kT\log\Delta\Xi[J_+,J_-],
\end{equation}
where
\begin{equation}
\label{gj1}
\Delta\Xi[J_+,J_-]=
\int D\Delta\rho_+\int D\Delta\rho_-
\exp\Big[-\beta \Big( \Delta\Omega^{MF}[\Delta\rho_+,\Delta\rho_-]
-\int_{\bf r}J_{\alpha}({\bf
r})\Delta\rho_{\alpha}({\bf r})\Big)\Big],
\end{equation}
and
\begin{equation}
\label{Xif1}
 \Delta\Omega^{MF}[\Delta\rho_+,\Delta\rho_-]=
\Omega^{MF}[\rho^*_{0+}+\Delta\rho_+,\rho^*_{0-}+\Delta\rho_-]-
\Omega^{MF}[\rho^*_{0+},\rho^*_{0-}].
\end{equation}
The last term in Eq.(\ref{gpf}) is the contribution to the grand
potential associated with fluctuations $ \Delta\rho_+,\Delta\rho_-$
around the most probable distributions, $\rho^*_{0+}({\bf r})$ and
$\rho^*_{0-}({\bf r})$. In the MF approximation this term is just
neglected, and the grand thermodynamic potential is approximated by
the minimum of $\Omega^{MF}[\rho_+,\rho_-]-\int_{\bf r}J_{\alpha}({\bf
r})\rho_{\alpha}({\bf r})$ with respect to $\rho_{\alpha}$ for fixed
$J_{\alpha}({\bf r})$. The average values of the local densities,
\begin{eqnarray}
\label{av}
\langle\rho_{\alpha}({\bf r})\rangle= \bar\rho_{\alpha}({\bf r})=
 \rho^*_{0\alpha}({\bf r})
+
\Delta\Xi[J_+,J_-]^{-1}\int D\Delta\rho_+\int D\Delta\rho_-
\Delta\rho_{\alpha}({\bf r})\times\\ \nonumber
\exp\Big[-\beta \Big( \Delta\Omega^{MF}[\Delta\rho_+,\Delta\rho_-]
-\int_{\bf r}J_{\alpha}({\bf
r})\Delta\rho_{\alpha}({\bf r})
\Big)\Big],
\end{eqnarray}
in MF are identified with their most probable values $
\rho^*_{0\alpha}({\bf r})$, i.e. the fluctuation contribution in
 Eq.(\ref{av}) is neglected. Beyond MF $\langle\rho_{\alpha}({\bf
 r})\rangle$ and $ \rho^*_{0\alpha}({\bf r})$ may differ from each
 other, as found in the case of the RPM~\cite{ciach:04:1,ciach:06:2}.
 In the context of the PM the role of the fluctuation contribution in
 Eqs.(\ref{gpf}) and (\ref{av}) will be studied in the future work, by
 following the Brazovskii approach~\cite{brazovskii:75:0} applied to
 the RPM in Ref.~\cite{ciach:06:1}.
 
In the MF approximation the generating functionals for the vertex and
the correlation functions, $F[\rho_+,\rho_-]$ and $-\Omega[J_+,J_-]$,
reduce to $\Omega^{MF}[\rho_+,\rho_-]$ and
$-\Big(\Omega^{MF}[\rho_+,\rho_-]-\int_{\bf r}J_{\alpha}({\bf
r})\rho_{\alpha}({\bf r})\Big)$ respectively. For $J_{\alpha}=0$ the
two-point correlation  and vertex functions,   $G^0_{\alpha\beta}$ and  $C^0_{\alpha\beta}$ respectively, in the above version of MF
are identical to the corresponding functions in the Gaussian
approximation. In the Gaussian approximation $\Omega^{MF}$ in the
Boltzmann factor (\ref{bolfac}) is approximated by Eq.(\ref{OmF}) with
the last term neglected. For $\Omega^{MF}=\Omega_2$ the exact
two-point correlation functions, $G^0_{\alpha\beta}$, are related to
$C^0_{\alpha\beta}$ given in Eq.(\ref{Calphabeta}) by the OZ equations
analogous to Eq.(\ref{OZ}).
\subsection{Phase transitions, spinodals and pretransitional effects}
Let us focus on the FT predictions for the phase diagram in the case
of $J_{\alpha}=0$. The stable (metastable) phases correspond to the
global (local) minimum of $F[\bar\rho_+,\bar\rho_-]$ (Eqs.(\ref{F})
and (\ref{gpf1})). Two minima corresponding to different forms of
$\bar\rho_{\alpha}({\bf r})$ are of equal depth at the coexistence between the
corresponding phases.  Let us assume that the global minimum of
$F[\bar\rho_+,\bar\rho_-]$ corresponds to $\rho^*_{0\alpha}({\bf r})$,
which deviates significantly from the constant function
$\rho^*_{0\alpha}$, corresponding to another extremum of
$F[\bar\rho_+,\bar\rho_-]$. Crystalline phases are characterized by
such density distributions. The minmum of $F$ corresponding to such
nonuniform distributions can be found only when $\Omega_{int}$ in
Eqs.(\ref{OmF}) and (\ref{gpf1}) is included, because the Gaussian
part $\Omega_2$ describes correctly only infinitesimal deviations from
the uniform distributions of ions $\rho^*_{0\alpha}({\bf r})$. $\Omega_{int}$ is given in terms of
the three- and higher-order vertex
functions in the MF approximation. Thus, the two-point vertex (or correlation) functions are
not sufficient to find the transitions to the ordered phases; it is
essential to include in Eq.(\ref{gpf1}) the higher-order vertex
functions as well.

 Let us consider the boundary of the global (local) stability of the uniform
 phase. The extremum of $F$
 changes character at the corresponding continuous-transition
 (spinodal) line.  At the high and at the low-$T^*$ side of this
  line the homogeneous phase corresponds to a global (local) minimum,
 and to a saddle point of $F$ respectively. More precisely, for given
 $\zeta$ the boundary of stability of the uniform phase is given by
 the highest $T^*$, for which $\det\tilde C_{\alpha\beta}(k)=0$
 ($\delta F/\delta\bar\rho_{\alpha}\delta\bar\rho_{\beta}$ is not
 positive definite) for some $k=k_b$.  In MF the boundary of the
 stability of the uniform phase is obtained
 with $\tilde C_{\alpha\beta}(k)$ approximated by $\tilde
 C^0_{\alpha\beta}(k)$ (Eq.(\ref{Calphabeta})). From the earlier
 studies~\cite{brazovskii:75:0,ciach:06:1} it follows that when $\det
 \tilde C^0_{\alpha\beta}(k)=0$ for $k\ne 0$, the fluctuation
 contribution to $F$ in Eq.(\ref{gpf1}) is comparable to
 $\Omega^{MF}$. Hence, the difference between $\tilde
 C_{\alpha\beta}(k)$ and $\tilde C^0_{\alpha\beta}(k)$ may be
 comparable to $\tilde C^0_{\alpha\beta}(k)$. In such a case the
 fluctuation contributions to Eqs.~(\ref{gpf1})-(\ref{av}) lead to
 significant shifts of the phase boundaries. In addition, if a
 continuous transition to a phase with periodic densities is found in
 MF, a fluctuation-induced first-order transition is
 expected at significantly lower temperatures~\cite{brazovskii:75:0,ciach:06:1} when the fluctuation
 contribution is included.

 The analysis of the stability of $
\Omega^{MF}[\rho_+,\rho_-]$ cannot give information on the actual instability of the disordered phase, but may give information
on the pretransitional effects associated with weak long-range
ordering of the instantaneous states.  Let us compare the probability
that an instantaneous state $\Delta\rho_+({\bf x})\ne 0,
\Delta\rho_-({\bf x})\ne 0$ occurs, with the probability of finding
the homogeneous instantaneous state $\Delta\rho_+=\Delta\rho_-= 0$.
According to Eqs.(\ref{bolfac}) and (\ref{OmF}), the probability ratio
is
\begin{equation}
\label{pratio}
\frac{p[\Delta\rho_+,\Delta\rho_-]}{p[0,0]}=
e^{-\beta(\Omega_{2}+\Omega_{int})}.
\end{equation}
Let the smaller eigenvalue of $\tilde
C^0_{\alpha\beta}(k)$ in Eq.(\ref{OmF2}) be denoted by $\tilde
C^0_{\phi\phi}(k)$.  
 The  eigenmode $\tilde\phi({\bf k})$ corresponding to $k=k_0$
  can be written as a linear combination of the functions
\begin{equation}
\label{eig}
\tilde g_i({\bf k}|k_0)=
\frac{(2\pi)^d}{\sqrt 2}\Big(w\delta({\bf k}-{\bf k}_{0i})+w^*
\delta({\bf k}+{\bf k}_{0i})\Big)
\end{equation}
with different directions of the vectors ${\bf k}_{0i}$, where $|{\bf
k}_{0i}|=k_0$ and $ww^*=1$. Let us focus on the instantaneous structure that has
the form $\tilde \phi({\bf k})=\Phi_i\tilde g_i({\bf k}|k_0)$, and the
amplitude of the other eigenmode of $\tilde C^0_{\alpha\beta}(k_0)$,
$\tilde\eta({\bf k})$, vanishes. From Eq.(\ref{OmF2}) we have
$\beta\Omega_2=V\sum_i\Phi_i^2\tilde C_{\phi\phi}(k_0)/2$, where $V$
is the volume of the system. Then Eq.(\ref{pratio}) takes the form
\begin{eqnarray}
\label{pratio_as}
\frac{\bar p[\Phi_i\tilde g_i({\bf k}|k_0),0]}{p[0,0]}=\exp\Bigg[-
\frac{\sum_i\Phi_i^2V}{2}\Big(\tilde C_{\phi\phi}(k_0)+O(\Phi_j)\Big)
\Bigg].
\end{eqnarray}
  In the above $\bar p$ is the probability (\ref{pratio}) as a
  function of the new variables, $\tilde\phi({\bf k})$ and
  $\tilde\eta({\bf k})$.

The sign of the term $O(\Phi_j)$ depends on the form of $\Omega_{int}$
in Eq.(\ref{OmF}), and on the form of $\tilde \phi({\bf k})$. Here we
limit ourselves to infinitesimal amplitudes $\Phi_i$.  When the term
$O(\Phi_i)$ in (\ref{pratio_as}) is negligible, the probability ratio
is larger or smaller from unity for $\tilde C_{\phi\phi}(k_0)<0$ or
for $\tilde C_{\phi\phi}(k_0)>0$ respectively. For given $\zeta$, the solution of
$\tilde C_{\phi\phi}(k_0)=0$ yields the temperature $T^*(\zeta;k_0)$,
and $\tilde C_{\phi\phi}(k_0)>0$ is equivalent to
$T^*>T^*(\zeta;k_0)$ (see (\ref{Calphabeta}) and below). 
  The most probable eigenmode corresponds to $k_b$ such that
$\tilde C^0_{\alpha\beta}(k)$ assumes a minimum at $k=k_b$.  Hence,
the instantaneous uniform state occurs with a higher probability than
the periodic instantaneous state with infinitesimal amplitude and {\it
any wavelength}  for $\tilde C_{\phi\phi}(k_b)>0$, i.e. for
$T^*>T^*(\zeta;k_b)$.  When for given $\zeta$ temperature decreases
from $T^*=T^*(\zeta;k_b)$, the range of $k$ such that
$\bar p[\Phi_i\tilde g_i({\bf k}|k),0]/p[0,0]>1$ 
increases. This means that the population of the
instantaneous periodic states that occur more frequently than the
instantaneous uniform state increases.  Ordering of the significant
fraction of the instantaneous states can be recognized as a
pretransitional effect associated with the transition to the
corresponding ordered phase.  The line in the phase diagram
corresponding to $\tilde C^0_{\phi\phi}(k_b)=0$ can be identified with
the structural line, because different instantaneous states dominate
on different sides of this line.
 Note that the structural line coincides precisely with the boundary of
stability of the uniform phase  in the MF approximation (MF spinodal).

 Near the minimum of $\tilde C^0_{\alpha\beta}(k)$ the probability
 ratio in Eq.(\ref{pratio_as}) varies slowly with $k$. Further studies
 are required to determine the probability that different wave
 packages corresponding to $k\approx k_b$ are thermally excited, and
 to verify if the wave packages are related to clusters of ions in 
 real space.  

We should note that the large-amplitude periodic states
 may be more probable than the uniform state even for $\tilde
 C^0_{\alpha\beta}(k_b)>0$, if the term $O(\Phi_j)$ in
 Eq.(\ref{pratio_as}) is negative. Note also that in the above studies
 we assumed that the instantaneous states did not include the
 eigenmode $\eta$ associated with the larger eigenvalue of $\tilde
 C^0_{\alpha\beta}(k)$. Interesting information on pretransitional
 effects can be gained from studies of the effect of coupling between
 the two eigenmodes in $\Omega_{int}$. In particular, the probabilites
 of the eigenmodes $\eta$, and $\phi$, given by $\int D\phi
 \bar p[\phi,\eta]$ and $\int D\eta \bar p[\phi,\eta]$ respectively,
 are of considerable interest. In the case of the
 RPM~\cite{ciach:01:0} the charge-charge correlations play a role analogous
 to short-range interactions between the
 ions~\cite{stell:95:0,ciach:01:0,patsahan:04:0}, and at low volume fractions $\int D\phi \bar
 p[\phi,\eta]$ turnes out to be important for the phase
 separation~\cite{ciach:01:0,ciach:06:1}. The above questions go
 beyond the scope of this work.
\section{Eigenmodes and MF boundary of stability of the disordered phase 
in the general case of arbitrary $\delta$ and $\nu$}
\subsection{Eigenmodes}
Our purpose here is to find the instantaneous states 
that may lead to instabilities of the disordered phase for arbitrary
$\delta$ and $\nu$ in the MF. In order to find the spinodal lines in
the MF (structural lines), it is sufficient to analyze $\beta\Omega_2$
(see Eqs.(\ref{OmF2}) and (\ref{pratio})).  $\tilde C^0_{\alpha\beta}(k)$
(Eq.(\ref{Calphabeta})) can be diagonalized, and $\Omega_2$ can be
written in the form
\begin{equation}
\label{Omei}
\beta\Omega_2=\frac{1}{2}\int \frac {d\bf k}{(2\pi)^2} \Big[
\tilde\phi({\bf k})\tilde C_{\phi\phi}^0(k)\tilde\phi(-{\bf k})+
\tilde\eta({\bf k})\tilde C_{\eta\eta}^0(k)\tilde\eta(-{\bf k})\Big].
\end{equation}
 Both
 $\tilde C_{\phi\phi}^0(k)$
and $\tilde C_{\eta\eta}^0(k)$ depend on the wavenumber $k$ in a nontrivial way,
\begin{equation}
\label{Cphi}
\tilde C_{\phi\phi}^0(k)=
\frac{\tilde C^0_{++}(k)+\tilde C^0_{--}(k)
-sign(\tilde C^0_{+-}(k)) {\cal B}(k)}{2}
\end{equation}
and
\begin{equation}
\label{Ceta}
\tilde C_{\eta\eta}^0(k)=
\frac{\tilde C^0_{++}(k)+\tilde C^0_{--}(k)
+sign(\tilde C^0_{+-}(k)) {\cal B}(k)}{2},
\end{equation}
where
\begin{equation}
\label{calB}
{\cal B}(k)=\sqrt{{\cal A}^2+4\tilde C^0_{+-}(k)^2},
\end{equation}
and
\begin{equation}
\label{calA}
{\cal A}(k)=sign(\tilde C^0_{+-}(k))
\big[\tilde C^0_{--}(k)-\tilde C^0_{++}(k)\big].
\end{equation}
In the above $\tilde C^0_{\alpha\beta}$ are given in Eq.(\ref{Calphabeta}),
$a_{\alpha\beta}$ are given in Appendix A and $\tilde
V_{\alpha\beta}(k)$ are given in Eqs.(\ref{Vf++})-(\ref{Vf+-}).

In the general case the corresponding eigenmodes have the forms
\begin{equation}
\label{v2}
\tilde\phi({\bf k})=\tilde a(k)\Delta\tilde \rho_{+}^*({\bf k})
-\tilde b(k)\Delta\tilde \rho_{-}^*({\bf k}),
\end{equation}
\begin{equation}
\label{v1}
\tilde\eta({\bf k})=\tilde b(k)\Delta\tilde \rho_{+}^*({\bf k})
+\tilde a(k)\Delta\tilde \rho_{-}^*({\bf k}),
\end{equation}
with
\begin{equation}
\label{a}
\tilde a(k)=\Bigg[\frac{{\cal A}(k)+
{\cal B}(k)}{2{\cal B}(k)}\Bigg]^{1/2},
\end{equation}
\begin{equation}
\label{b}
\tilde b(k)=\Bigg[\frac{-{\cal A}(k)
+{\cal B}(k)}{2{\cal B}(k)}\Bigg]^{1/2}.
\end{equation}
 The eigenmodes represent two order-parameter (OP) fields, and in
 principle either one of them may lead to instability of the
 disordered phase.

On the MF level the FT analogs (\ref{OZ}) of the OZ equations are
decoupled in the eigenbasis of the vertex functions, i.e. 
\begin{equation}
\label{OZMF}
\tilde
G^0_{\phi\phi}(k)=1/\tilde C^0_{\phi\phi}(k)\hskip1cm {\rm and} \hskip1cm\tilde
G^0_{\eta\eta}(k)=1/\tilde C^0_{\eta\eta}(k).
\end{equation}
 Note that in the fully
symmetrical case ($\nu=\delta=0$), $\tilde C^0_{--}(k)=\tilde
C^0_{++}(k)$, and $\tilde a(k)=\tilde b(k)=1/\sqrt 2$, hence $\phi$
and $\eta$ are proportional to the charge- and the number density
respectively. Moreover, Eqs.(\ref{Cphi}) and (\ref{Ceta}) reduce to
$\tilde C^0_{\phi\phi}(k)= (\tilde C^0_{++}(k)+\tilde
C^0_{--}(k)-2\tilde C^0_{+-}(k))/2$ and $\tilde C^0_{\eta\eta}(k)=(
\tilde C^0_{++}(k)+\tilde C^0_{--}(k)+2\tilde C^0_{+-}(k))/2$, 
representing the charge-charge and the density-density vertex function
respectively.  When any
asymmetry is present and $\tilde C^0_{--}(k)\ne \tilde C^0_{++}(k)$,
$C^0_{\phi\phi}$ and $C^0_{\eta\eta}$ that satisfy the decoupled OZ
equations differ from the charge- and number density vertex functions
respectively, in agreement with Ref.\cite{stell:95:0}.
 
\subsection{Dominant order parameters}
 Let us focus on thermodynamic conditions such that the uniform
 distributions $\rho^*_{0\alpha}$ correspond to the minimum of
 $\Omega^{MF}$, and consider the two OP fields with equal wavelengths
 and equal infinitesimal amplitudes. The field leading to a smaller
 increase of $\Omega^{MF}$ is thermally excited with a higher
 probability (\ref{bolfac}), and such OP dominates over the
 other one.  Which one of the two infinitesimal fields,
 $\tilde\phi({\bf k})$ or $\tilde\eta({\bf k})$ dominates, depends on
 which function, $\tilde C_{\phi\phi}(k)$ or $\tilde C_{\eta\eta}(k)$,
 is smaller (see (\ref{Omei})). This, in turn, depends on $sign(\tilde
 C_{+-}(k))$ (see (\ref{Cphi}) and (\ref{Ceta})). From
 Eqs.(\ref{Calphabeta}) and (\ref{Vf+-}) we obtain the line
\begin{equation}
\label{domflu}
T^*=\frac{4\pi\cos k}{k^2}a_{+-}^{-1}
\end{equation}
 separating the phase space $(\zeta,T^*)$ into the high-temperature
 part where the OP $\tilde\phi({\bf k})$ dominates, and the
 low-temperature part where the other eigenmode dominates. In the
 above $a_{+-}$ is a function of $\zeta$ (or $s$) given in Appendix
 A. Note that for different wavenumbers $k$ Eq.(\ref{domflu}) yields
 quite different lines. In particular, for long-wavelengths
  $k\to 0$, corresponding to phase separation into two
 uniform phases, from (\ref{domflu}) we find that $\tilde\eta({\bf
 0})$ dominates for all temperatures $T^*<\infty$.  We can thus
 conclude that if the phase separation (stable or metastable) occurs,
 it is induced by the field $\eta$.  On the other hand, for $k\ge
 \pi/2$, corresponding to the periodic ordering of ions with the
 wavelength $2\pi/k\le 4$ (in $\sigma_{+-}$ units), the OP
 $\tilde\phi({\bf k})$ dominates for all temperatures $T^*>0$. In
 principle, the uniform system may become unstable with respect to
 the eigenmodes with $0<k<\pi/2$ as well. As we show later, this is
 indeed the case for very large asymmetry.  In such a system $\tilde\phi({\bf
 k})$ or $\tilde\eta({\bf k})$ dominates for temperatures higher or
 lower than the temperature given in Eq.(\ref{domflu}) respectively.
\subsection{The MF spinodal lines}

Let us first analyze the stability of the uniform phase with respect
to long-wavelength deviations from homogeneous distributions of ions in the MF approximation. For
$k\to 0$ Eqs.(\ref{Ceta}) and (\ref{Cphi}) assume the asymptotic forms
\begin{eqnarray}
\label{Cnetaasy}
\tilde C^0_{\eta\eta}(k)=
\frac{a_{++}Z^{-1}+a_{--}Z+2a_{+-}-4\pi\beta^*\delta^2}
{Z+Z^{-1}}+O(k^2),
\end{eqnarray}
and
\begin{eqnarray}
\label{Cphiasy}
\tilde C^0_{\phi\phi}(k)=\frac{4\pi\beta^*(Z+Z^{-1})}{k^2}+O(1).
\end{eqnarray}
$\tilde C^0_{\phi\phi}(k)$ diverges for $k\to 0$. When $\tilde
C^0_{\phi\phi}(0)=\infty$, then the probability $p\propto
\exp(-\beta\Omega^{MF}[\phi,\eta])$ of thermally exciting any nonzero
$\tilde\phi(0)$ vanishes (see (\ref{OmF}) and (\ref{Omei})), as
required by the global charge neutrality. $\tilde C^0_{\eta\eta}(k)$ can
vanish for $k=0$ when
\begin{eqnarray}
\label{spin0}
T^*=\frac{4\pi\delta^2}{a_{++}Z^{-1}+a_{--}Z+2a_{+-}}.
\end{eqnarray}
The above equation describes the MF spinodal line associated with the
phase separation into two uniform phases characterized by different
values of the OP $\eta$.

 The MF
instability with respect to the $k$-mode  is given by 
\begin{eqnarray}
\label{inst}
\det \tilde C^0_{\alpha\beta}(k)=
\tilde C^0_{\eta\eta}(k)\tilde C^0_{\phi\phi}(k)=0,
\end{eqnarray}
 where
\begin{eqnarray}
\label{1.2}
\tilde C^0_{\eta\eta}(k)\tilde C^0_{\phi\phi}(k)=
(a_{++}+\beta \tilde V_{++}(k))(a_{--}+
\beta \tilde V_{--}(k))-(a_{+-}+\beta \tilde V_{+-}(k))^2=\\ \nonumber
-\frac{\sin^2(k\delta)}{k^4}\big(4\pi\beta^*)^2+
\frac{b(k,s,\delta,\nu)}{k^2} 4\pi\beta^*+d(s,\delta,\nu),
\end{eqnarray}
and the functions $b(k,s,\delta,\nu)$ and
$d(s,\delta,\nu)>0$ are given in Appendix B.

Note that there is a single positive solution of Eq.(\ref{inst}) for
$\beta^*$ for all values of the remaining parameters.  The wavector at
the MF spinodal (structural line) is determined from
\begin{eqnarray}
\label{wvs}
\partial (\tilde C^0_{\eta\eta}(k)\tilde C^0_{\phi\phi}(k))/\partial k=0.
\end{eqnarray}
The solutions of Eqs.~(\ref{inst}) and (\ref{wvs}) give the spinodal
line and the wavenumber of the dominant instantaneous state. 

\subsection{Nature of the eigenmodes for different $\delta$ and $\nu$}
It is instructive
to discuss the nature of the eigenmodes that may drive the system out
of the disordered state.  The
equations (\ref{v2}) and (\ref{v1}) can be easily inverted to give
\begin{equation}
\label{v1in}
\Delta\tilde \rho_{+}^*({\bf k})=\tilde a(k)\tilde\phi({\bf k})
+\tilde b(k)\tilde\eta({\bf k}),
\end{equation}
\begin{equation}
\label{v2in}
\Delta\tilde \rho_{-}^*({\bf k})=-\tilde b(k)\tilde\phi({\bf k})
+\tilde a(k)\tilde\eta({\bf k}).
\end{equation}
 The local  charge
density, 
\begin{equation}
\label{q}
q({\bf r})=Z\Delta\rho_+^*({\bf
r})-\Delta\rho_-^*({\bf r})
\end{equation}
(in $|e_-|\sigma_{+-}^{-3}$ units), and
the local deviation of the number density of ions from the most probable
value (in $\sigma_{+-}^{-3}$ units), 
\begin{equation}
\label{roav}
\Delta\rho({\bf r})=\Delta\rho_+^*({\bf
r})+\Delta\rho_-^*({\bf r}),
\end{equation}
are related to the two OP's by
\begin{eqnarray}
\label{chargenumber}
\tilde q({\bf k})= \Big(Z\tilde a(k)+\tilde b(k)\Big)\tilde\phi({\bf k})+
\Big(Z\tilde b(k)-\tilde a(k)\Big)\tilde\eta({\bf k})\\ \nonumber
\Delta\tilde\rho({\bf k})=\Big(\tilde a(k)-\tilde b(k)\Big)\tilde\phi({\bf k})+
\Big(\tilde a(k)+\tilde b(k)\Big)\tilde\eta({\bf k}).
\end{eqnarray}
The inverse relations are
\begin{equation}
\label{inv1}
\tilde\phi({\bf k})=\frac{1}{Z+1}
\Big[(\tilde a(k)+\tilde b(k))\tilde q({\bf k})+
(\tilde a(k)-Z\tilde b(k))\Delta\tilde\rho({\bf k})\Big]
\end{equation}
\begin{equation}
\label{inv2}
\tilde\eta({\bf k})=\frac{1}{Z+1}\Big[(\tilde b(k)-
\tilde a(k))\tilde q({\bf k})+(Z\tilde a(k)+
\tilde b(k))\Delta\tilde\rho({\bf k})\Big].
\end{equation}

 On the low-temperature side of the structural line the dominant instantaneous states consist of the
 charge- and the number-density waves with the wavnumber $k_b$
 and the amplitudes that satisfy the relation
\begin{equation}
\label{fqd}
\Delta\tilde\rho({\bf
 k}_b)=R\tilde q({\bf k}_b).
\end{equation}
As seen from Eqs.(\ref{chargenumber})-(\ref{inv2}), the instability of
$\Omega^{MF}$ associated with $\tilde\phi({\bf k}_b)$ (i.e. with
$\tilde\eta({\bf k}_b)=0$), corresponds to
\begin{equation}
\label{fqdphi}
R=\frac{(\tilde a(k_b)-\tilde
 b(k_b))}{(Z\tilde a(k_b)+\tilde b(k_b))}.
\end{equation}
 For $\tilde a(k_b)=\tilde b(k_b)$, i.e. in the RPM limit, $R=0$, and 
 $\tilde\phi(k_b)\propto\tilde q({\bf k}_b) $.  When the instability of
$\Omega^{MF}$
 is induced by $\tilde\eta({\bf k}_b)$ and the other OP vanishes, then
\begin{equation}
\label{fqdeta}
R= \frac{(\tilde a(k_b)+\tilde b(k_b))}{(Z\tilde
 b(k_b)-\tilde a(k_b))}. 
\end{equation}
  For
 $Z\tilde b(k_b)=\tilde a(k_b)$ we have $1/R=0$, and 
 $\tilde\eta({\bf k}_b)\propto\Delta\tilde\rho({\bf k}_b)$.

For $R\ll 1$ the charge-density waves dominate over the number-density
waves in the dominant eigenmode. In such a case the regions of the
excess positive charge are followed by regions of excess negative
charge. The number densities in the charged regions are
comparable. For $R\gg 1$ the number-density waves dominate,
i.e.  regions with excess number-density are followed by 
regions containing a smaller number of ions. Both the dense  and the dilute regions
 are nearly charge-neutral for $R\gg 1$.

 In real-space representation the OP fields are
given by convolutions,
\begin{equation}
\label{convolution}
\phi({\bf r})=\int d{\bf r}_1 
[\Delta\rho^*_+({\bf r}_1)a({\bf r}-{\bf r}_1)-
\Delta\rho^*_-({\bf r}_1)b({\bf r}-{\bf r}_1)],
\end{equation}
with analogous expression for the field $\eta({\bf r})$, where $a({\bf
r})$ and $b({\bf r})$ are inverse Fourier transforms of the functions
defined in Eq.(\ref{a}) and (\ref{b}).

\section{Case of equal sizes}
\subsection{MF spinodal and the wavevector of the most probable instantaneous state}
The phase separation (\ref{spin0}) occurs only for different sizes of
the ionic species ($\delta\ne 0$).  In the size-symmetric case the
disordered phase is unstable only with respect to periodic ordering in
MF.  Let us consider the periodic ordering, for which the MF spinodal is
given by Eqs.~(\ref{inst}) and (\ref{wvs}). For $\delta=0$ we obtain
\begin{eqnarray}
\label{e=0a}
a_{\alpha\alpha}=\frac{1}{\rho^*_{0\alpha}}+a_{+-}
\end{eqnarray}
and 
\begin{eqnarray}
\label{e=0apm}
a_{+-}=\frac{\pi(4-s)(2+s^2)}{(1-s)^4}.
\end{eqnarray}
The boundary of stability of the disordered
phase and the associated wavevector reduce to the form
\begin{eqnarray}
\label{e=0}
T^*_R(s)=-\frac{24\cos k_R}{k_R^{2}}s,\hskip1cm \tan k_R=-\frac{2}{k_R}.
\end{eqnarray}
The above form is identical to the one found for the RPM within the
same approach\cite{ciach:00:0,ciach:05:0}. The subscript $R$ indicates
that the MF spinodal and the wavenumber refer to the RPM. The charge
asymmetry has no effect on the boundary of stability of the disordered
phase on the MF level of our theory, as long as the sizes are the
same. This result agrees with MF predictions of other
theories~\cite{patsahan_mryglod:06,netz:99:0,caillol:05:0,caillol:04:0}.
 Beyond MF this  property does not 
persist~\cite{patsahan_mryglod:06,caillol:05:0,caillol:04:0} due to
the coupling between the two OP's in $\Omega_{int}$.
\subsection{Eigenmodes}
In this case we obtain the following  form of Eq.(\ref{calA})
\begin{eqnarray}
{\cal A}=-sign\tilde C_{+-}(k)\frac{4\nu}{1-\nu^2}
\Bigg(\frac{\pi}{6s}+\frac{4\pi\beta^*\cos k}{k^2}\Bigg).
\end{eqnarray}
At the MF spinodal line (\ref{e=0})  ${\cal A}=0$, and in turn $\tilde
a(k_R)=\tilde b(k_R)$ (see Eqs.(\ref{a}) and (\ref{b})), thus
\begin{equation}
\label{pp}
\tilde\phi({\bf k}_R)=\frac{1}{\sqrt 2}\Bigg(\Delta\tilde\rho_+({\bf k}_R)
-\Delta\tilde\rho_-({\bf k}_R)\Bigg)
\end{equation}
\begin{equation}
\label{pp1}
\tilde\eta({\bf k}_R)=\frac{1}{\sqrt 2}\Bigg(\Delta\tilde\rho_+({\bf k}_R)
+\Delta\tilde\rho_-({\bf k}_R)\Bigg).
\end{equation}
These relations are the same as in the RPM, although for $Z\ne 1$
 $\tilde\phi({\bf k})$ differs from the charge-density amplitude (\ref{q}).
Beyond the spinodal line ${\cal A}\ne 0$, and the eigenmodes contain
different proportions of $ \Delta\tilde\rho_+$ and $
\Delta\tilde\rho_-$.
\section{The MF spinodal associated with 
the gas-liquid phase separation}
\subsection{Eigenmodes in the case of $k\to 0$}
Let us discuss the nature of the eigenmodes for $k\to 0$.
From 
Eqs.(\ref{a}) and (\ref{b}) we find  for
arbitrary $\delta$ and $\nu$
\begin{equation}
\label{kto0}
\tilde a(0)= \frac{Z}{\sqrt{1+Z^2}},\hskip1cm \tilde b(0)=  \frac{1}{\sqrt{1+Z^2}}.
\end{equation}
In the long-wavelength limit the amplitudes of the eigenmodes assume
the forms
\begin{equation}
\label{kto0eta}
\tilde\eta(0)=\frac{1}{\sqrt{1+Z^2}}
(\Delta\tilde\rho^*_+(0)+Z\Delta\tilde\rho^*_-(0))
\end{equation}
\begin{equation}
\label{kto0phi}
\tilde\phi(0)=\frac{1}{\sqrt{1+Z^2}}
(Z\Delta\tilde\rho^*_+(0)-\Delta\tilde\rho^*_-(0)).
\end{equation}
  Note that from Eqs.(\ref{kto0phi}) and (\ref{q}) it follows that 
   $\tilde\phi(0)\propto \tilde q(0)$.  The
  charge-neutrality condition (\ref{neut}) is equivalent to
  $\tilde\phi(0)=0$.
 For
  $\tilde\phi(0)=0$ we obtain
\begin{equation}
\label{etq}
\tilde\eta(0)=\frac{\sqrt{1+Z^2}}{1+Z}\Delta\tilde\rho(0).
\end{equation}
In sec.IIIB we have shown that for $k\to 0$ the dominant eigenmode is
$\tilde\eta(0)$. Eq.(\ref{etq}) shows that $\tilde\eta(0)\propto
\Delta\tilde\rho(0)$. Hence, we find the usual phase separation into
ion-dilute and ion-dense phases for all values of the asymmetry
parameters, as expected.
\subsection{MF Spinodal lines}
In our MF approximation the spinodal line for the gas-liquid
separation assumes a single maximum $T^*_{c}(\zeta_{c})$ at the
critical point (cp) for any pair of $\delta\ne 0$ and $\nu$. The MF
spinodal lines are shown in Figs.1,2 and 3 for $\delta=0.2$,
$\delta=0,6$ and $\delta=0.9$ respectively, for several values of
$\nu$. The spinodal lines assume the characteristic asymmetric 
shape as in the RPM~\cite{stell:95:0,fisher:94:0}.
\begin{figure}
\includegraphics[scale=0.45]{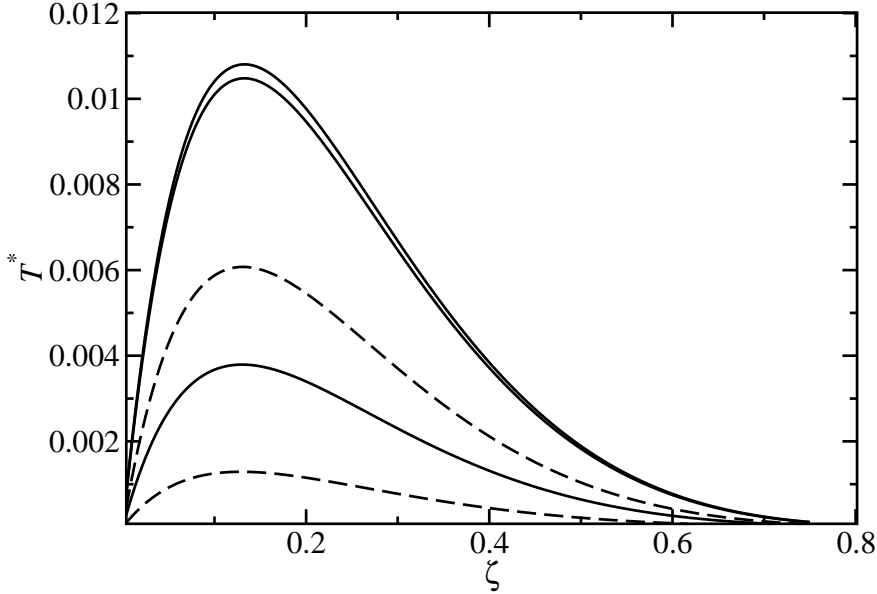}
\caption{MF spinodal lines for the gas-liquid separation for 
$\delta=0.2$ ($\lambda=3/2$). The solid and the dashed lines
correspond to $\nu\ge 0$ and $\nu<0$ respectively.
 From the top to
the bottom, the solid lines correspond to $\nu=0.5$, $\nu=0$
and $\nu=0.9$. The upper and the lower dashed line correspond to
$\nu=-0.5$ and $\nu=-0.9$ respectively.}
\end{figure}
\begin{figure}
\includegraphics[scale=0.45]{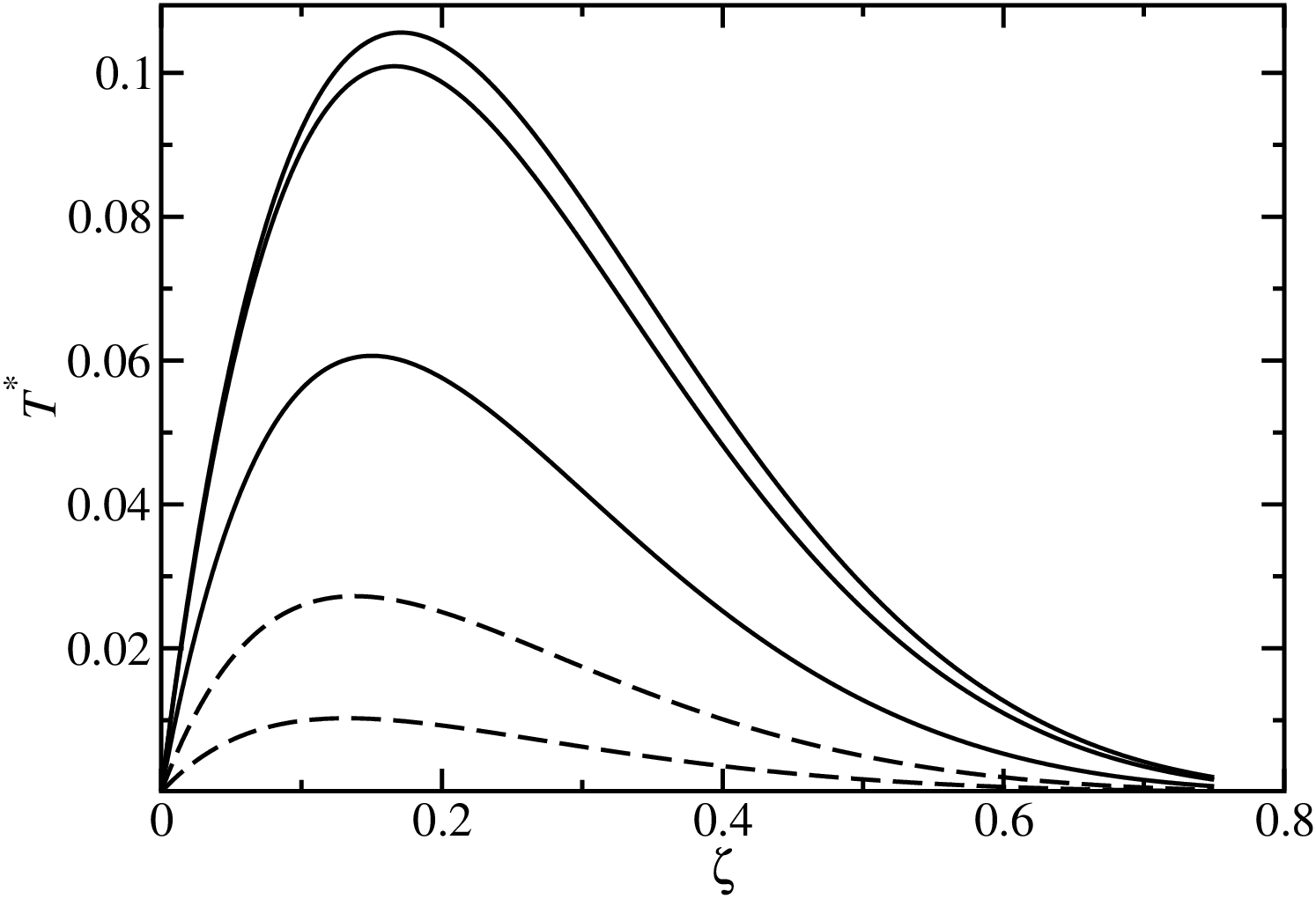}
\caption{MF spinodal lines for the gas-liquid separation for 
$\delta=0.6$ ($\lambda=4$). The solid and the dashed lines
correspond to $\nu\ge 0$ and $\nu<0$ respectlively.
 From the top to
the bottom, the solid lines correspond to $\nu=0.9$, $\nu=0.5$
and $\nu=0$. The upper and the lower dashed line correspond to
$\nu=-0.5$ and $\nu=-0.8$ respectively.}
\end{figure}
\begin{figure}
\includegraphics[scale=0.45]{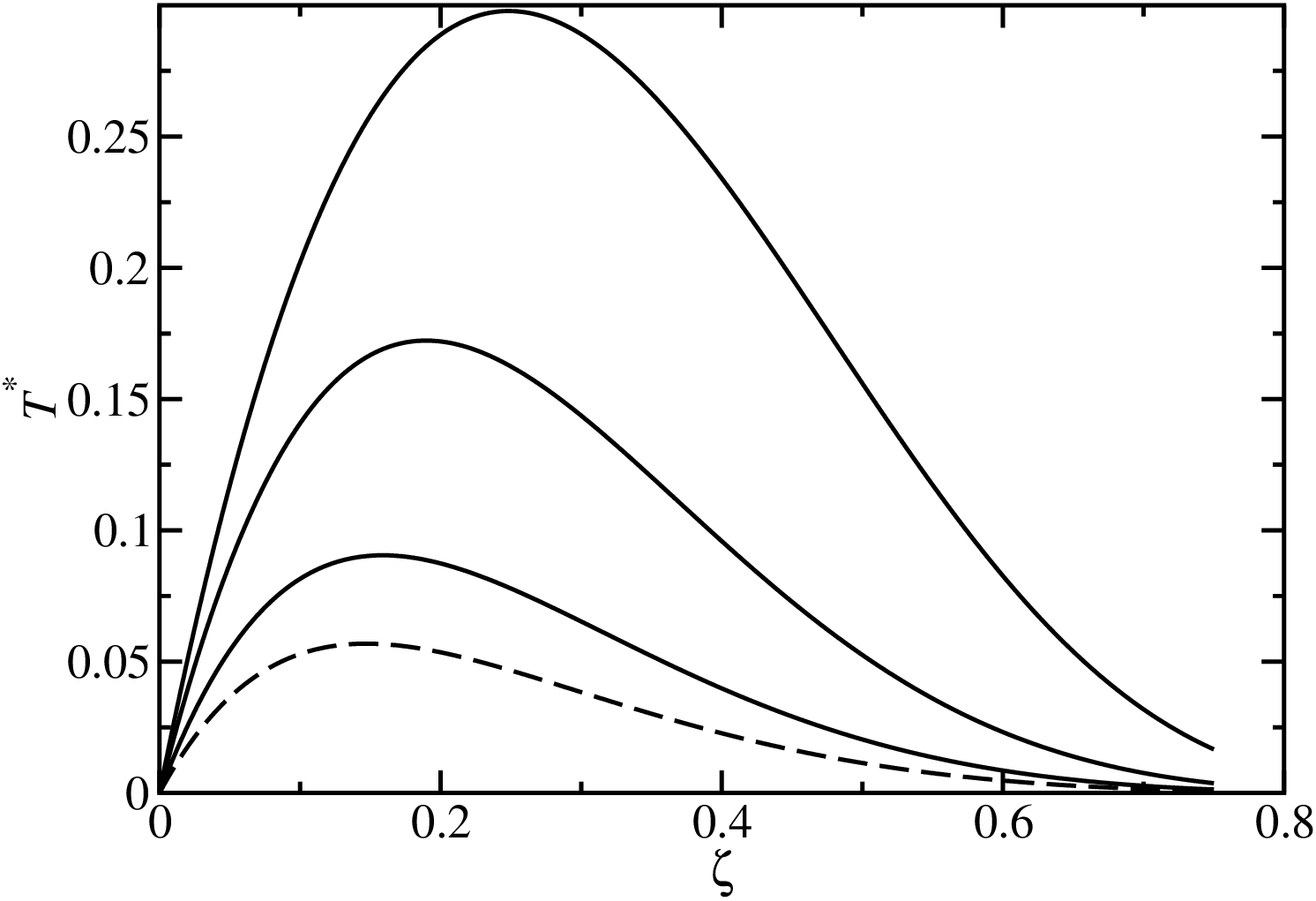}
\caption{MF spinodal lines for the gas-liquid separation for 
$\delta=0.9$ ($\lambda=19$). The solid and the dashed lines
correspond to $\nu\ge 0$ and $\nu<0$ respectlively. 
From the top to
the bottom, the solid lines correspond to $\nu=0.9$, $\nu=0.5$
and $\nu=0.0$. The  dashed line corresponds to
$\nu=-0.3$.}
\end{figure}
 The critical temperature and the volume fraction, $T^*_{c}$ and
 $\zeta_{c}$ respectively, are shown in Fig.4 as functions of the size
 asymmetry for several charge ratios. For a given charge ratio the
 dependence of $\zeta_{c}$ on the size asymmetry agrees qualitatively
 with the major trends found in Monte Carlo (MC) simulations
\cite{romero:00:0,cheong:03:0,yan:01:0,yan:02:0,yan:02:1}.
 Namely, $\zeta_{c}$ is a convex function of $\delta$, and its
 value is of the same order of magnitude as in
 Ref.~\cite{cheong:03:0}. 
 In contrast,
 $T^*_{c}$ is a convex function of $\delta$ for fixed $\nu$,
 whereas in MC a concave function was obtained
\cite{romero:00:0,cheong:03:0,yan:01:0,yan:02:0}. 
For large asymmetries the value of $T^*_c$ is overestimated, whereas for $\delta=0$ we have  $T^*_c=0$ in our MF
approximation.  On the other hand, the dependence of $T^*_{c}$ on
$\nu$ for fixed $\delta$ agrees qualitatively with MC
results~\cite{cheong:03:0}. Namely, for $\delta<0$ the $T^*_{c}$
decreases with increasing charge asymmetry, and for $\delta>0$ the lines
$T^*_c(\delta)$ corresponding to different charge ratios intersect
each other (Fig.4 (upper panel) here and Fig.9 in Ref.~\cite{cheong:03:0}). In the case
of $\zeta_{c}$, a correct dependence on the charge asymmetry is found
for $\delta>0$, whereas for $\delta<0$ our results do not agree
with (partial) results obtained in Ref.~\cite{cheong:03:0}).

 In the case of $\delta=1$ and $\nu\to 1$ ($\lambda=\infty$,
 $Z\to\infty$), and for $\zeta=O(Z^0)$, Eq.(\ref{spin0}) assumes the
 simple form
\begin{equation}
\label{sp1}
T^*=3\zeta(1-\zeta)^2.
\end{equation}
The critical-point temperature, $T^*_{c}=4/9$, and density,
$\zeta_{c}=1/3$, are both large compared to the corresponding values in the RPM, in contrast to
the simulation results.
  Thus, some of the trends found in simulations
are correctly predicted by the MF theory, while some other trends are
not.  Note, however
that beyond MF coupling between the two OP's may have a significant
effect on the spinodal. In particular, beyond MF $T^*_c>0$ is found for
$\delta=0$~\cite{ciach:00:0,ciach:05:0} as a result of
the coupling between the two OP's in $\Omega_{int}$ (see
Eq.(\ref{OmFint})).  Significant effect of the coupling between the
two OP's on the spinodal should be expected also for $\delta\ne0$
and $\nu\ne 0$.  Indeed, a qualitatively correct dependence of the
location of the cp on $\nu$ is found for $\delta=0$ in the
CV approach
\cite{patsahan_mryglod:06}, beyond MF. As mentioned in sec.IIA, the CV approach and our mesoscopic
theory are  closely related~\cite{patsahan:06:3}.
 Correct
trends are also found when ion pairing is explicitly
taken into account~\cite{aqua:05:0,zhou:05:0}.
\begin{figure}
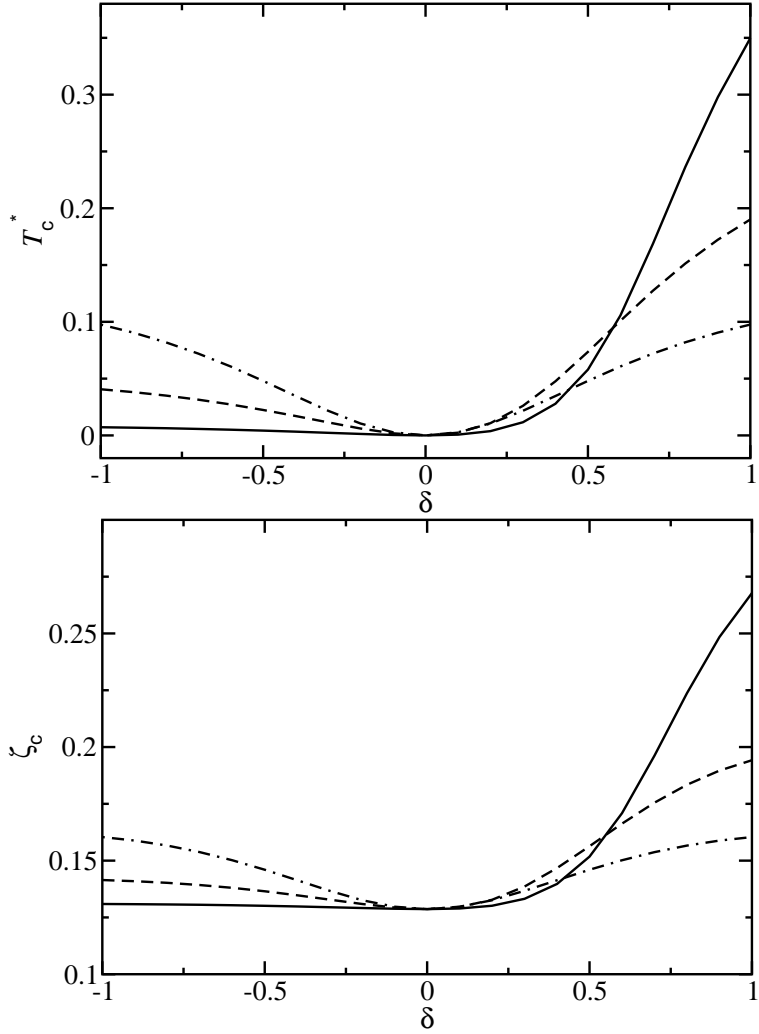

\includegraphics[scale=0.4]{fig4a.eps}
\hskip3cm 
\includegraphics[scale=0.4]{fig4b.eps}
\caption{Critical temperature (top) and critical volume fraction 
(bottom) as functions of the size asymmetry $\delta$, for three values
of the charge asymmetry $\nu$.  Solid, dashes and dash-dotted lines
correspond to $\nu=0.9,0.5$ and $0$ respectively.  }
\end{figure}

\section{The MF spinodal associated with periodic ordering of ions} 
The structural line (MF spinodal) associated with the periodic
ordering of the instantaneous states, as well as the most probable
periodic structures depend qualitatively on $\delta$ and $\nu$. The
key role is played by the size asymmetry. Three regimes of the size
asymmetry can be distinguished, although there are no sharp
boundaries, but rather smooth crossovers between them. The approximate
ranges of the first two regimes are $\delta< 0.4$ ( $\lambda<2.3$),
and $0.4<\delta<0.9$ ($2.3<\lambda<19$). The qualitative dependence on
$\nu$ is found in the second regime only. Third regime corresponds to
the asymptotic case of $\delta,\nu\to 1$, i.e. to extreme charge- and
size asymmetry. In the other (unphysical) extreme, $\delta,\nu\to -1$, the periodic
ordering is suppressed.  Detailed description of each case is given
in the following subsections.
\subsection{Small size asymmetry}

\subsubsection{MF spinodal lines and  wavevectors in the dominant instantaneous states}
The MF spinodal in the case of the small size-asymmetry $\delta<0.4$ $
(\lambda<2.3)$ is qualitatively the same as in the RPM, i.e. along the
spinodal $T^*$ is a monotonically increasing function of $\zeta$
(Fig.5).  Somewhat surprising is the fact that for larger
charge-asymmetry the ordering occurs at higher temperatures. It is not
clear whether this tendency persists beyond MF. The wavenumber
characterizing the period of the oscillations of the OP field in the
ordered phase is $k_b>\pi/2$, and its value is almost independent of
$\zeta$ (see Fig.6). The size of the regions of increased or depleted
densities of the two ionic species, $\sim
\pi/k_b<(\sigma_++\sigma_-)$, is comparable to the size of the ions,
and the ordered phase should be identified with a standard 'hard'
ionic crystal. According to the discussion in sec.III.B, the relevant
OP field is $\tilde\phi(k_b)$ for the whole temperature range. The
amplitudes of the number- and the charge-density waves in the dominant
instantaneous state are related according to
Eqs.(\ref{fqd})-(\ref{fqdphi}). The amplitude ratios $R$ for
$\delta=0.2$ are shown in Fig.7.

\begin{figure}
\includegraphics[scale=0.5]{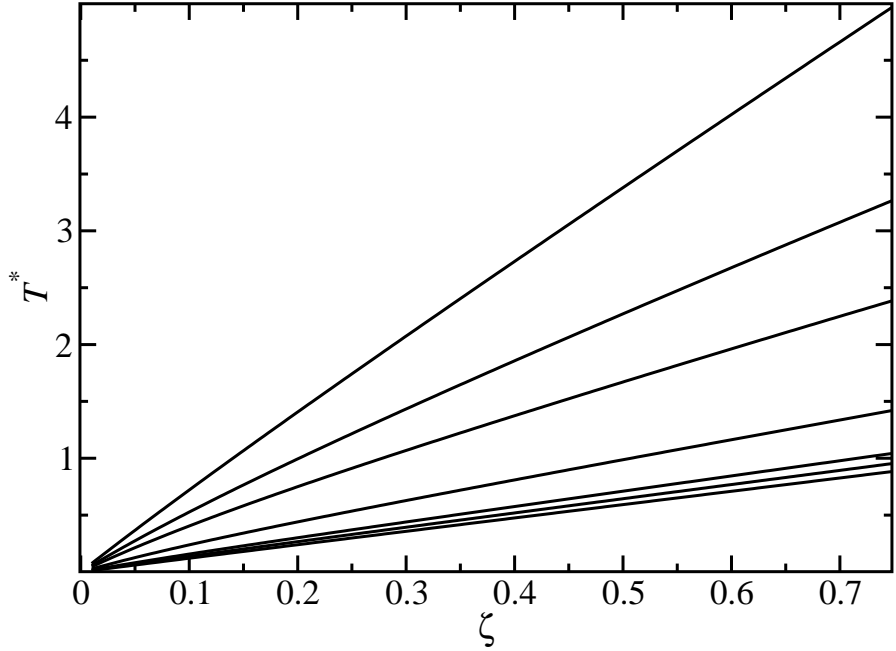}
\caption{MF spinodal lines for the transition to the ordered phase
 for $\delta=0.2$. From the top to the bottom
 lines $\nu=0.9$, $\nu=0.7$, $\nu=0.5$
 $\nu=0.0$, $\nu=-0.5$, $\nu=-0.7$ and
 $\nu=-0.9$. Temperature $T^*$ and the volume fraction of ions
 $\zeta$ are in dimensionless reduced units defined in Eqs.(\ref{def})
 and (\ref{zeta}) respectively.}
\end{figure}
\begin{figure}   
\includegraphics[scale=0.5]{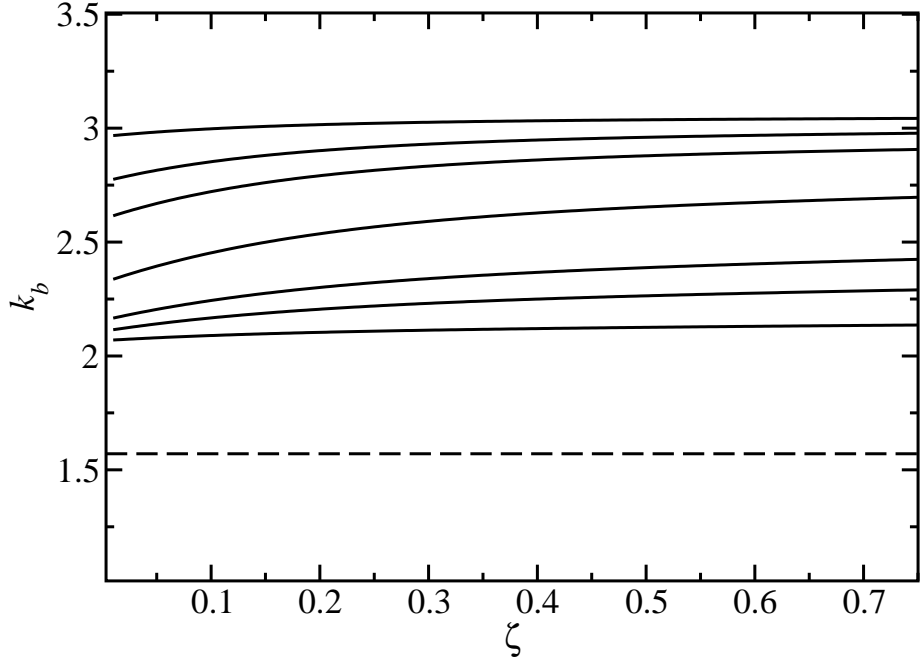}
\caption{The wavenumber $k_b$ corresponding to the ordering of ions 
 along the MF spinodal lines shown in Fig.5 for $\delta=0.2$. From
 the bottom to the top lines $\nu=0.9$, $\nu=0.7$, $\nu=0.5$
 $\nu=0.0$, $\nu=-0.5$, $\nu=-0.7$ and $\nu=-0.9$. $k_b$
 is in $\sigma_{+-}^{-1}$ units and $\zeta$ is the volume fraction of
 ions.}
\end{figure}

 The above results indicate that the behavior of the PM for
 $\delta<0.4$ is qualitatively the same as in the RPM. By analogy with
 the RPM, we expect that the transition to the ionic crystal is
 fluctuation-induced first-order, and occurs at significantly lower
 $T^*$ and higher $\zeta$ compared to the structural
 line~\cite{ciach:06:2}.
\begin{figure}
\includegraphics[scale=0.5]{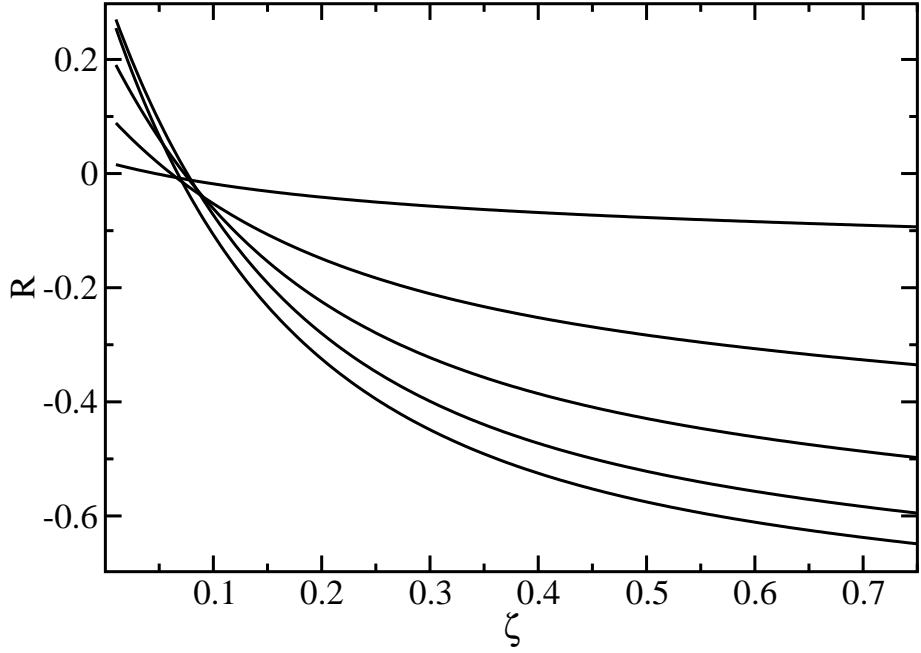}
\caption{The ratio $R$ in the relation
  (\ref{fqd}) between the amplitudes of the number- and charge-density
  waves in the dominant eigenmode, along the MF spinodal lines shown in
  Fig.5 for $\delta=0.2$. From the top to the bottom lines (on the
  right) $\nu=0.9$, $\nu=0.5$, $\nu=0.0$, $\nu=-0.5$ and
  $\nu=-0.9$. Both $R$ and the volume fraction of ions $\zeta$ are
  dimensionless. Note that $|R|<1$, indicating periodic ordering into
  positively and negatively charged regions of the spacial extent
  $\pi/k_b$ (in $\sigma_{+-}$ units), where $k_b$ is shown in Fig.6. }
\end{figure}
\subsubsection{Eigenmodes for very small asymmetries}
Let us determine the behavior of the eigenmodes 
in the RPM limit.  For  $\delta,\nu\to 0$ we find from Eq.(\ref{calA}) 
that ${\cal A}\to 0$, and 
\begin{eqnarray}
\label{ab}
\tilde a(k)\simeq_{\delta,\nu\to 0}\frac{1}{\sqrt 2}
\big(1+y\big) +O(\delta^2,\delta\nu,\nu^2)\\
\nonumber
\tilde b(k)\simeq_{\delta,\nu\to 0}\frac{1}{\sqrt 2}
\big(1-y\big) +O(\delta^2,\delta\nu,\nu^2)
\end{eqnarray}
where 
\begin{eqnarray}
\label{y}
y=\frac{{\cal A}}{2{\cal B}}=O(\delta)+O(\nu).
\end{eqnarray}
The eigenmodes (\ref{v2}) and (\ref{v1})  take the forms
\begin{eqnarray}
\label{eigen_small}
\tilde\phi({\bf k})=\frac{1}{\sqrt 2}
\Bigg[\Delta\tilde\rho_+({\bf k})-\Delta\tilde\rho_-({\bf k})+
y
\Big(\Delta\tilde\rho_+({\bf k})+\Delta\tilde\rho_-({\bf k})\Big)\Bigg]
+O(\delta^2,\delta\nu,\nu^2) 
\\
\nonumber
=\frac{1}{\sqrt 2}\Big[(1-\nu)\tilde q({\bf k})+
(y-\nu)\Delta\tilde\rho({\bf k})\Big]+
O(\delta^2,\delta\nu,\nu^2)
\end{eqnarray}
and
\begin{eqnarray}
\label{eigen_small2}
\tilde\eta({\bf k})=\frac{1}{\sqrt 2}
\Bigg[\Delta\tilde\rho_+({\bf k})+\Delta\tilde\rho_-({\bf k})-
y
\Big(\Delta\tilde\rho_+({\bf k})-\Delta\tilde\rho_-({\bf k})\Big)\Bigg]+
O(\delta^2,\delta\nu,\nu^2)\\
\nonumber
=\frac{1}{\sqrt 2}\Big[\Delta\tilde\rho({\bf k})-y\tilde q({\bf k})\Big] 
+O(\delta^2,\delta\nu,\nu^2).
\end{eqnarray}
  For infinitesimal asymmetry parameters, $\Delta\rho$ and $q$ yield
  infinitesimal contribution to $\phi$ and to $\eta$ respectively. The
  OZ equations are thus decoupled for the eigenmodes $\phi$ and $\eta$ that very weakly
  deviate from the charge- and the number densities respectively.
This result agrees with the predictions of
  Ref.~\cite{stell:95:0}.

\subsection{Moderate and large size asymmetry}
For $\delta>0.4$ ($\lambda>2.3$) the structural line associated with the
periodic ordering of the instantaneous states assumes a single
maximum for $\zeta\approx 0.15$. Unlike  the case of $\delta<0.4$,
 at higher volume fractions the periodic ordering is less favorable. 
 For the nonmonotonic structural lines two cases can be
distinguished, according to a different behavior of the wavenumber of
the dominant instantaneous states $k_b$.

\subsubsection{$k_b>\pi/2$ }
Such large wavenumbers are found for the moderate size asymmetry,
$0.4<\delta<0.67$, in the case of $\delta\nu>0$ (larger charge at the
larger ion). For $\delta=0.6$ and $\nu>0$ the structural lines and the
corresponding wavenumbers $k_b$ are shown in Figs.8 and 9
respectively.  Except for very small $\zeta$, the wavenumber is
$k_b>\pi/2$, and increases with increasing $\zeta$. Increasing $k_b$
indicates decreasing period of oscillations of $\rho_{\alpha}({\bf
x})$, in agreement with expected smaller nearest-neighbor distances
for larger density. For $k_b>\pi/2$ the dominant field is $\phi$
(sec.IIIB). The behavior of $R$, shown in Fig.10, is similar as in
the case $\delta<0.4$.

The analysis of the cluster formation in Ref.\cite{spohr:02:0} was
performed for this range of parameters. Namely, the case of
$\delta=0.639$, and $\nu=0.85$ (monovalent), $\nu=0.7$ (divalent) and
$\nu=0.6$ (trivalent counterions) was examined in detail. The volume
fractions and temperatures were $(\zeta, T^*)= (0.0185,0.157),
(0.0174,0.079)$ and $(0.017,0.0525)$ for mono-, di- and trivalent
counterions respectively. All cases correspond to the low-$T^*$ side
of the structural line, where the periodic instantaneous structures
dominate.  Only for trivalent counterions $k_b<\pi/2$ and the relevant
OP is $\eta$, but this fact is not sufficient to explain the
differences found in the short-range ordering into clusters. Further
studies beyond the stability analysis of $\Omega^{MF}$ are required to
verify whether the coarse-grained description can explain the cluster
formation.
\begin{figure}
\includegraphics[scale=0.45]{fig8.eps}
\caption{MF spinodal lines of the transition to the ordered phase
 for $\delta=0.6$ and $\nu\ge 0$. From the top to the bottom
 lines  $\nu=0.9$, $\nu=0.7$, $\nu=0.5$ and
 $\nu=0.0$. Temperature $T^*$ and the volume fraction of ions
 $\zeta$ are in dimensionless reduced units defined in Eqs.(\ref{def})
 and (\ref{zeta}) respectively.}
\end{figure}
\begin{figure}
\includegraphics[scale=0.45]{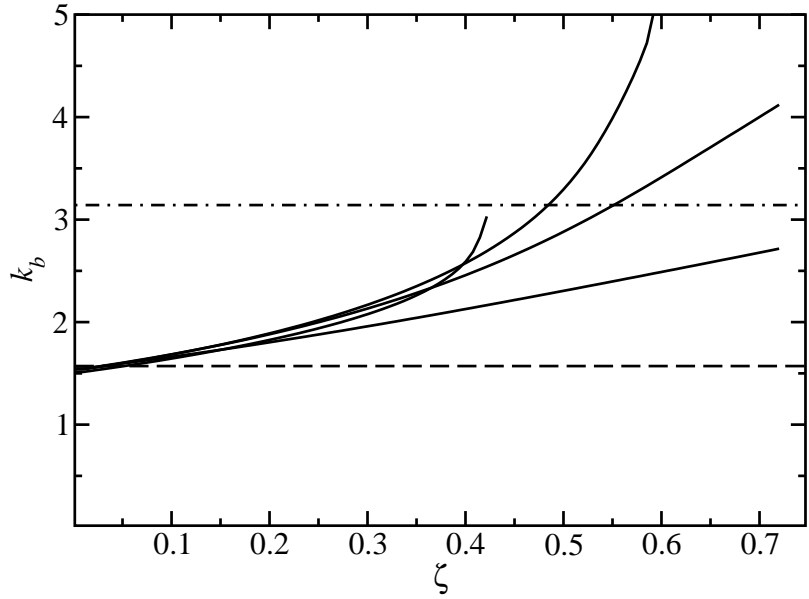}
\caption{The wavenumber $k_b$ corresponding to the ordering of ions 
 along the MF spinodal lines shown in Fig.8 (for $\delta=0.6$ and
 $\nu\ge 0$). From the bottom to the top lines $\nu=0.9$,
 $\nu=0.7$, $\nu=0.5$ and $\nu=0.0$. $k_b$ is in
 $\sigma_{+-}^{-1}$ units and $\zeta$ is the volume fraction of
 ions. The dashed and dash-dotted lines correspond to $k_b=\pi/2 $ and
 $\pi$ respectively. Note that $k_b<\pi/2$
 for very small values of $\zeta$.}
\end{figure}

\begin{figure}
\includegraphics[scale=0.4]{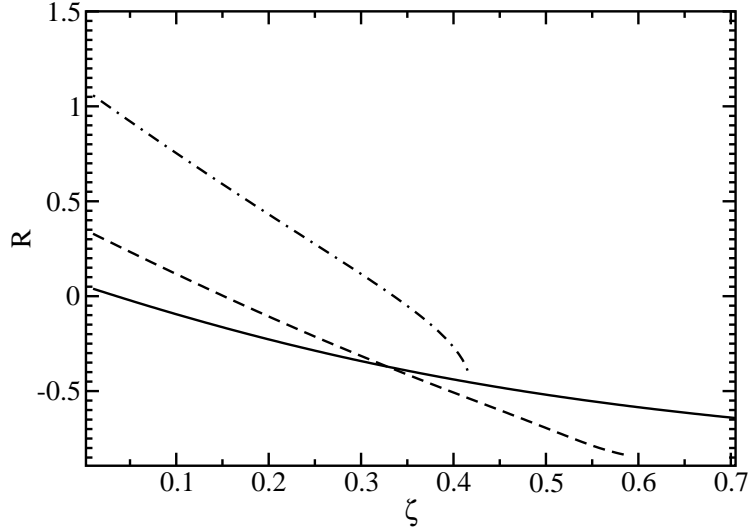}
\caption{The ratio $R$ (\ref{fqd}) between the amplitudes of the 
number- and charge-density waves in the dominant eigenmode for
$\delta=0.6$ and $\nu\ge 0$. The solid, dashed, and dash-dotted lines
correspond to $\nu=0.9$, $\nu=0.5$ and $\nu=0.0$ respectively.  Note
that $|R|<1$, and periodic ordering into positively and negatively
charged regions of a size $\pi/k_b$, with $k_b$ shown in Fig.9, takes
place.  Both $R$ and the volume fraction of ions $\zeta$ are
dimensionless. }
\end{figure}
\subsubsection{$k_b<\pi/2$}
  $k_b<\pi/2$ is found in the case of $0.4<\delta<0.67$  for $\nu<0$
 (larger charge at the smaller ion), as shown in Fig.11. 
\begin{figure}
\includegraphics[scale=0.45]{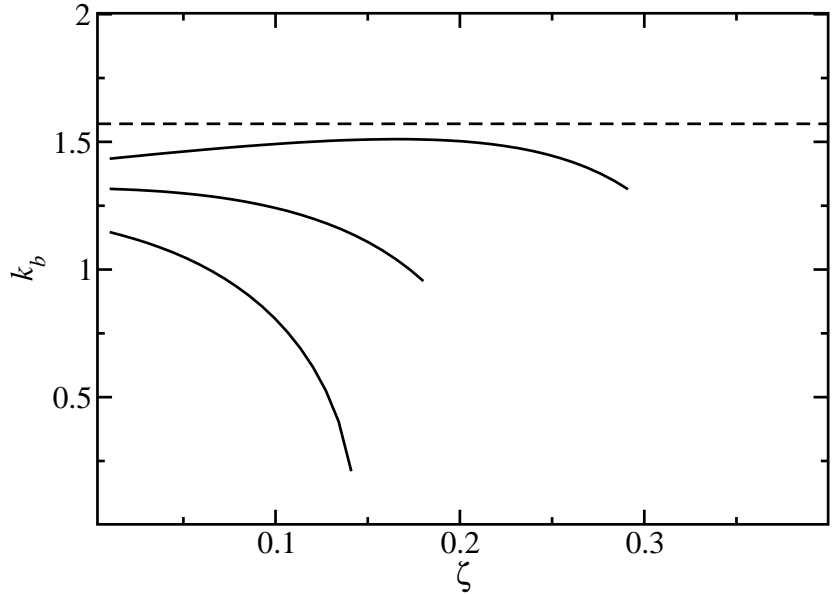}
\caption{The wavenumber $k_b$ corresponding to the ordering of ions 
 along the MF spinodal lines shown in Fig.12 (for $\delta=0.6$ and
 $\nu<0$). From the top to the bottom lines $\nu=-0.5$,
 $\nu=-0.7$ and $\nu=-0.8$. $k_b$ is in $\sigma_{+-}^{-1}$ units
 and $\zeta$ is the volume fraction of ions. At the dashed line 
$k_b=\pi/2$.}
\end{figure}
\begin{figure}
\includegraphics[scale=0.45]{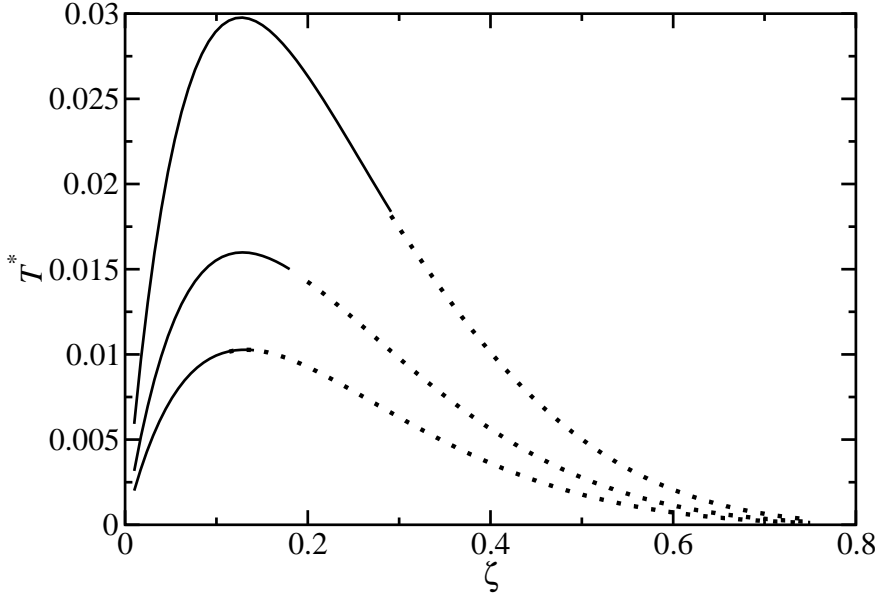}
\caption{Solid lines represent the MF spinodal lines of the 
transition to the ordered phase
 for $\delta=0.6$ and $\nu< 0$. The dotted lines are the MF
 spinodals corresponding to the gas-liquid phase separation associated
 with $k_b=0$, discussed in sec.V. From the top to the bottom lines
 $\nu=-0.5$, $\nu=-0.7$ and $\nu=-0.8$. Temperature $T^*$ and the volume 
fraction of ions
 $\zeta$ are in dimensionless reduced units defined in Eqs.(\ref{def})
 and (\ref{zeta}) respectively.}
\end{figure}

 For $\delta> 0.67$ ($\lambda>5$) the  wavenumber of the dominant eigenmode of $\tilde C_{\alpha\beta}^0(k_b)$ is
 $k_b<\pi/2$ for all values of $\nu$ (Fig.13).  The extent of the
 phase-space region corresponding to periodic ordering of the
 instantaneous states is similar to that of  the case of $0.4<\delta<0.67$
 (Figs.14 and 15).
\begin{figure}
\includegraphics[scale=0.45]{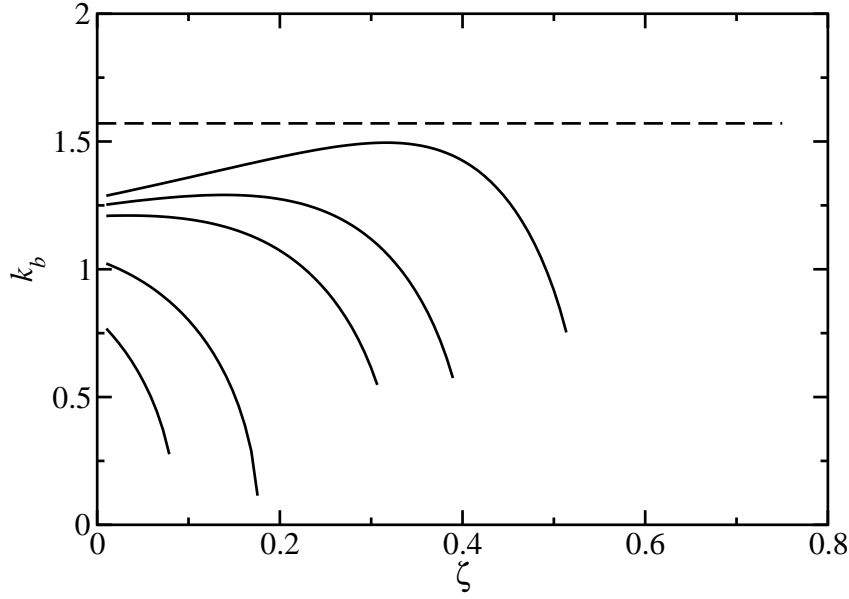}
\caption{The wavenumber $k_b$ corresponding to the ordering of ions 
 along the MF spinodal lines for $\delta=0.9$. From the top to the
 bottom lines $\nu=0.9$, $\nu=0.7$, $\nu=0.5$, $\nu=0.0$
 and $\nu=-03$. $k_b$ is in $\sigma_{+-}^{-1}$ units and $\zeta$ is
 the volume fraction of ions. The horizontal dashed line corresponds
 to $k_b=\pi/2 $.}
\end{figure}
\begin{figure}
\includegraphics[scale=0.45]{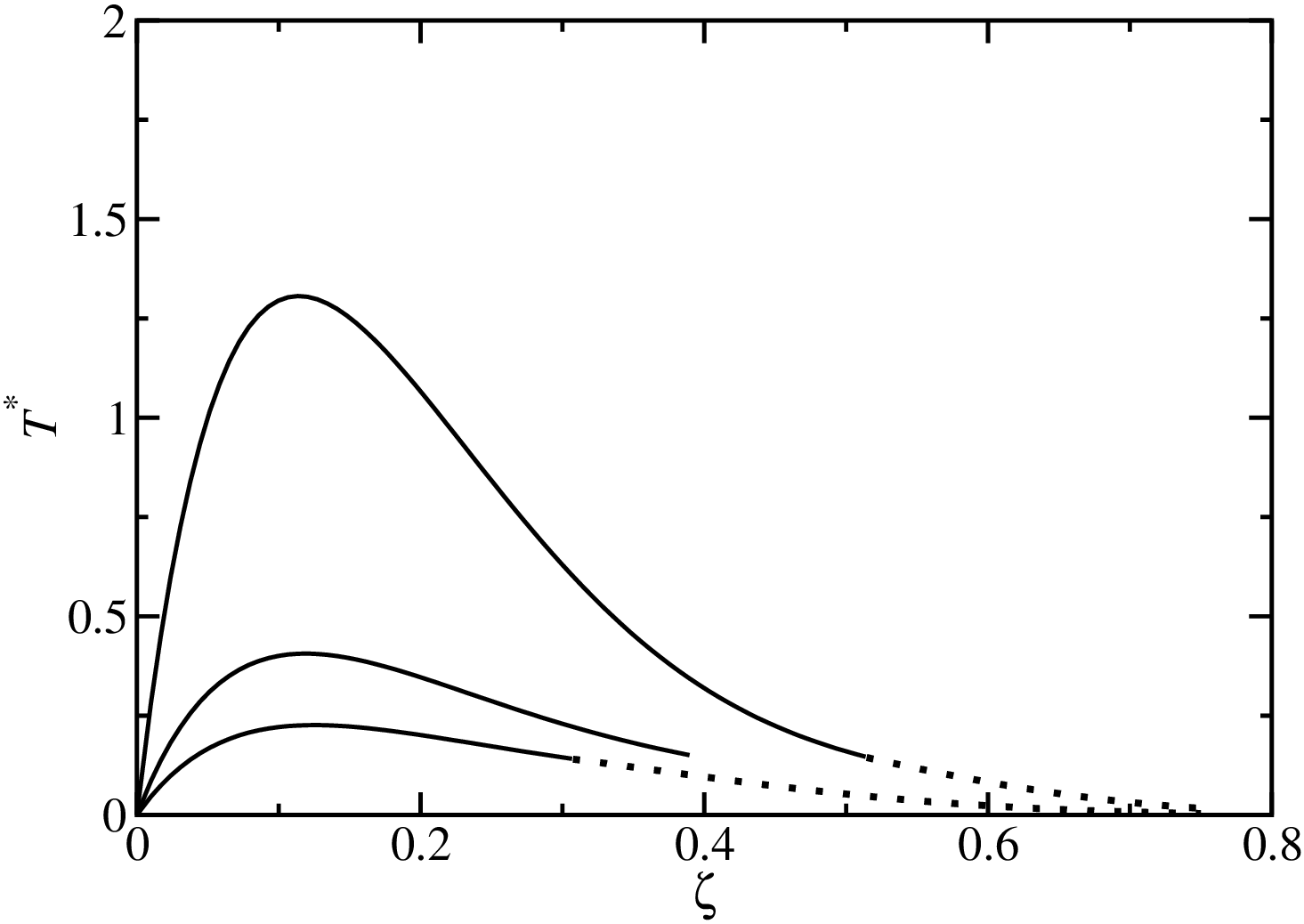}
\caption{MF spinodal lines (solid) for the transition to the
 ordered phase
 for $\delta=0.9$ and $\nu> 0$. Dotted 
line represents the MF spinodal
 line for the gas-liquid separation. From the top to the bottom lines
 $\nu=0.9$, $\nu=0.7$ and $\nu=0.5$. Temperature $T^*$ and
 the volume fraction of ions $\zeta$ are in dimensionless reduced
 units defined in Eqs.(\ref{def}) and (\ref{zeta}) respectively.}
\end{figure}

\begin{figure}
\includegraphics[scale=0.45]{fig15.eps}
\caption{MF spinodal lines (solid)  for the transition to the ordered 
phase for $\delta=0.9$ and $\nu\le 0$.  Dotted lines are the MF
spinodal lines for the gas-liquid separation. From the top to the
bottom lines $\nu=0.0$ and $\nu=-0.3$. Temperature $T^*$ and the
volume fraction of ions $\zeta$ are in dimensionless reduced units
defined in Eqs.(\ref{def}) and (\ref{zeta}) respectively.}
\end{figure}
\begin{figure}
\includegraphics[scale=0.4]{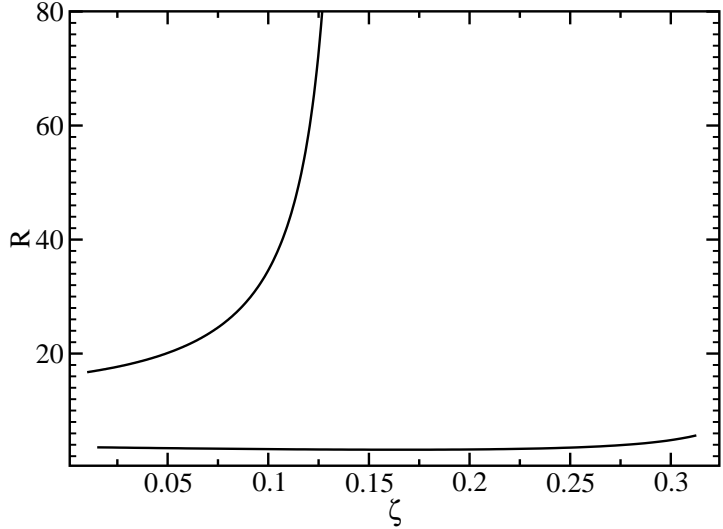}
\caption{The ratio $R$ (\ref{fqd}) between the amplitudes of the 
number- and charge-density waves in the dominant instantaneous
state for $\delta=0.6$ and $\nu<0$. The upper and the lower curve
correspond to $\nu=-0.8$ and $\nu=-0.5$. Note that $R>1$, and regions
of increased and depleted number density are formed. The size of these
regions is $\pi/k_b$, with $k_b$ shown in Fig.11. Both $R$ and the
volume fraction of ions $\zeta$ are dimensionless. In the case of the upper curve 
both the dense and the dilute regions in the dominant eigenmode are nearly charge neutral for $\zeta>0.1$.}
\end{figure}
When the difference between $\nu$ and $\delta$ is not too large, the
$k_b$ initially increases with increasing $\zeta$ (Fig.11 (upper
curve) and 13 (two upper curves)). When $k_b$ increases, the structural line and the MF
spinodal of the gas-liquid separations are well separated. The
amplitude ratio is $R=O(1)$ and varies slowly with $\zeta$. At a value
of $\zeta$ depending on $\delta$ and $\nu$, $k_b$ starts to decrease,
and at some point a rapid decrease to $k_b=0$ occurs.  At the
corresponding value of $\zeta=\zeta_L$ (Lifshitz point) the MF
spinodals merge together and become identical for
$\zeta>\zeta_L$ (Figs.12, 14 and 15). The  amplitude ratio $R$  (\ref{fqd})
increases
rapidly, when a rapid decrease to $k_b=0$ occurs.
For a large difference between the charge and the size asymmetry, the
$k_b$ decreases for the whole range of $\zeta$ (see the two lower
curves in Fig.11 and the three lower curves in Fig. 13). 

  For $R\gg 1$ the amplitude of the charge-density is much smaller
  than the amplitude of the number-density wave (sec.IIID). Thus, the
  ordering into alternating oppositely charged regions is
  suppressed. For $\nu<0$ a large charge at the small ion is compensated by a large
  number of large ions having small charges; therefore the unit cell
  of the ordered structure is large, and may contain voids.  In
crystals containing voids as parts of the structure the size of the
unit cell is significantly larger than the sum of radii of the two
ionic species. Such voids should
be distinguished from vacancies resulting from thermal fluctuations,
which are present in the ionic crystals at low densities.

In this case the favorable
  ordered structures are much more complex than in both, the RPM-like
  systems and the colloid-like systems.  Whether such complex
  structures may correspond to stable phases in the PM with moderate
  size- and charge asymmetry remains an open question, because the
  binodal associated with the periodic ordering may be preempted by
  the binodal associated with the phase separation. 

For a large difference between the charge and the size asymmetry, the
MF instability with respect to ordering of ions in periodic structures
occurs only for small values of $\zeta$ and $T^*$ (Figs. 12, 14, and 15).  For
$\nu\delta\to -1$ the ordering is suppressed entirely. An infinite
number of large ions is required to neutralize the infinite charge at
the small ion for $\nu=-1$. Clearly, formation of an ordered structure
with a finite period is not possible in this (unphysical) case.

Finally, we should note that at large volume fractions the effects of
ordering of hard spheres become important, and such effects are not
accounted for by this theory.

\begin{figure}
\includegraphics[scale=0.45]{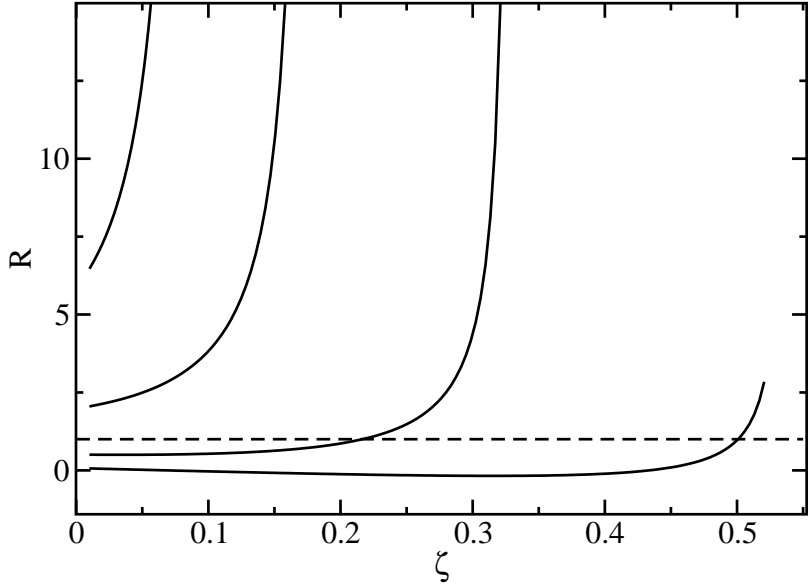}
\caption{The ratio $R$ (\ref{fqd}) between the amplitudes of the number- 
and charge-density deviations from the average values in the dominant
instantaneous state for $\delta=0.9$. From the bottom to the top lines
$\nu=0.9$, $0.5$, $0$, and $-0.3$. At the dashed line $R=1$. Both $R$
and the volume fraction of ions $\zeta$ are dimensionless.}
\end{figure}
\subsection{Case of very large asymmetry, $\delta\nu\to 1$}
The maximum of the MF spinodal $T^*(\zeta)$ increases when
$\nu\approx\delta\to 1$, and for $\nu\approx\delta>0.9$ the maximum
increases very rapidly. This increase of the phase-space region
where a majority of the instantaneous states are the periodically ordered
  can be observed by comparing
Figs.13, 18 and 19. In the latter plot the spinodal line and the
wavevector are shown for $Z=10^4$ ($\nu=0.9998$) and $\lambda=10^3$
($\delta=0.998$), corresponding to highly charged colloids with
counterions of a microscopic size, considered in many experimental studies.
\begin{figure}
\includegraphics[scale=0.45]{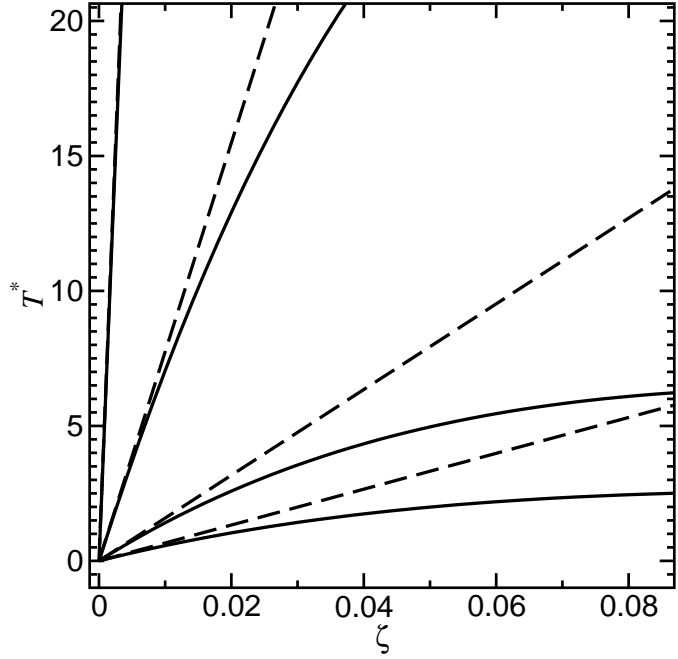}
\caption{The MF spinodal lines (solid) and the corresponding
 asymptotic behavior for $\nu= \delta\to 1$ (dashed). The latter has
 the form $T^*=12.3\zeta/(1-\delta^4)$ (Eq.38 in Ref.\cite{ciach:06:0}
 and Eq.(\ref{zetas}) here). From the bottom to the top lines
 $\delta=\nu= 0.95, 0.98,0.996$ and $0.9995$ ($\lambda=Z=39, 99, 499$
 and $3999$) respectively. In the last case the solid and the dashed
 lines are indistinguishable for the given range of $T^*$.  }
\end{figure}

\begin{figure}
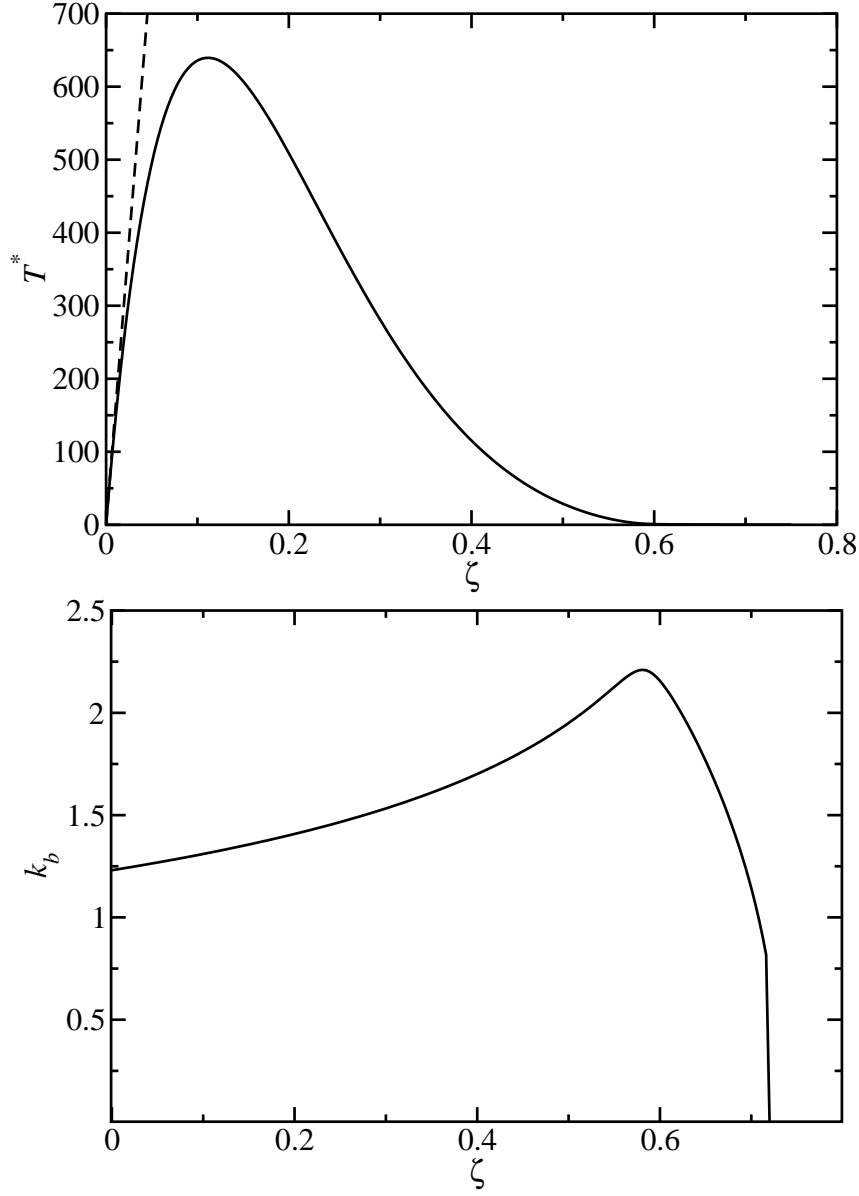

\includegraphics[scale=0.45]{fig19a.eps}\\
\includegraphics[scale=0.45]{fig19b.eps}
\caption{The MF spinodal line (top) and the corresponding wavevector
 of the dominant eigenmode (bottom) for $\delta=0.998$ and
 $\nu=0.9998$. The dashed line is the asymptotic result of
 Ref.\cite{ciach:06:0}. For $\zeta\ll 1$, $k_b\approx 1.23$ in
 agreement with the asymptotic result of Ref.\cite{ciach:06:0}. $T^*$
 is in the dimensionless units (Eq.(\ref{def})), and $k_b$ is in units
 of $\sigma_{+-}^{-1}\approx 2/\sigma_+$. }
\end{figure}

 The case of extreme asymmetry was studied in
 Ref.\cite{ciach:06:0} for $s=O(Z^0)$ ($\zeta=O(Z^{-1}))$, i.e. for
 infinite dilution of colloid particles. In the limiting case the
 spinodal assumes the asymptotic form \cite{ciach:06:0}
\begin{equation}
\label{asy}
T^*_C(\zeta)=-\frac{3\cos(2k_C)}{k_C^2}Z\zeta
\end{equation}
where
\begin{equation}
 \tan(2k_C)=-\frac{1}{k_C}.
\end{equation}
The subscript $C$ indicates that the MF spinodal and the corresponding
wavevector  correspond to the colloid limit, $\delta,\nu\to 1$, with the
infinite dilution of the colloid particles for $Z\to\infty$.

Note that $k_C=k_R/2<\pi/2$, and according to the discussion in
sec.IIIB the eigenmode $\phi$ is relevant on the high-$T^*$ side,
whereas the eigenmode $\eta$ is relevant on the low-$T^*$ side of the
line (\ref{domflu}). The asymptotic analysis in Ref.\cite{ciach:06:0}
indicates that the phase behavior is determined only by the field
$\Delta\rho_+$. Here we shall verify if the dominant eigenmode indeed
reduces to $\Delta\rho_+$.  For $Z\to\infty$ and for $\tilde
C_{++}(k)\ne 0$ we obtain the approximate form of Eq.(\ref{calB}),
\begin{eqnarray}
\label{Bcinfty1}
{\cal B}=|\tilde C_{++}(k)|-sign(\tilde C_{++}(k))\tilde C_{--}(k)+
\frac{2\tilde C_{+-}(k)^2}{|\tilde C_{++}(k)|}+O(Z^{-2}).
\end{eqnarray}
 In the limit $\delta=\nu=1$ the line (\ref{domflu}) assumes the form
 $T^*=\frac{3\cos k}{k^2}$. From (\ref{a}), (\ref{Bcinfty1}) and the
 above we obtain the asymptotic behavior of $\tilde a$ and $\tilde b$
 for $Z\to\infty$,
\begin{eqnarray}
\label{abcinfty2}
\tilde a(k)=\left\{
 \begin{array}{lllll}
1-O\Big(\frac{1}{Z^2}\Big) &\;\; \mbox{if} 
 &\;\;\; \mbox{$\frac{3\cos k}{k^2}<T^*<T^*_C(\zeta)$} &\;\; \mbox{or}
 &\;\;  \mbox{$T^*_C(\zeta)<T^*<\frac{3\cos k}{k^2}$}\\
O\Big(\frac{1}{Z}\Big) &\;\; \mbox{otherwise}  &\;\;\; 
\mbox{} 
&\;\; \mbox{}  &\;\; 
\mbox{},
\end{array}
 \right.
\end{eqnarray}
with $\tilde b(k)=O\Big(\frac{1}{Z}\Big)$ and $\tilde
b(k)=1-O\Big(\frac{1}{Z^2}\Big)$ in the first and in the second regime
respectively. The eigenmodes obtained from Eqs.(\ref{v2}) and (\ref{v1}) 
assume the forms
\begin{eqnarray}
\label{abcinfty3}
\tilde\phi({\bf k})\approx\left\{
 \begin{array}{lllll}
\tilde\Delta\rho_+({\bf k}) &\;\; \mbox{if} 
 &\;\;\; \mbox{$\frac{3\cos k}{k^2}<T^*<T^*_C(\zeta)$} &\;\; \mbox{or}
 &\;\;  \mbox{$T^*_C(\zeta)<T^*<\frac{3\cos k}{k^2},$}\\
\tilde\Delta\rho_- ({\bf k})&\;\; {\rm otherwise,}
 &\;\;\;\mbox{} &\;\; \mbox{}
 &\;\;  \mbox{}\\
\end{array}
 \right.
\end{eqnarray}
and 
\begin{eqnarray}
\label{abcinfty4}
\tilde\eta({\bf k})\approx\left\{
 \begin{array}{lllll}
\tilde\Delta\rho_-({\bf k}) &\;\; \mbox{if} 
 &\;\;\; \mbox{$\frac{3\cos k}{k^2}<T^*<T^*_C(\zeta)$} &\;\; \mbox{or}
 &\;\;  \mbox{$T^*_C(\zeta)<T^*<\frac{3\cos k}{k^2},$}\\
\tilde\Delta\rho_+({\bf k}) &\;\; {\rm otherwise},
 &\;\;\;\mbox{} &\;\; \mbox{}
 &\;\;  \mbox{}\\
\end{array}
 \right.
\end{eqnarray}
where the neglected contributions are $O(1/Z)$.   The dominant eigenmodes for different parts
of the $(\zeta,T^*)$ phase diagram, obtained from
Eqs.(\ref{abcinfty3}) and (\ref{abcinfty4}), are shown in Fig.20.
Below the MF spinodal line, i.e. for $T^*\le T^*_C(\zeta)$, each dominant
OP reduces to $\Delta\tilde\rho_+(k_C)$. The asymptotic analysis of
Ref.~\cite{ciach:06:0} is thus fully consistent with the present, more
complete approach.

\begin{figure}
\includegraphics[scale=0.45]{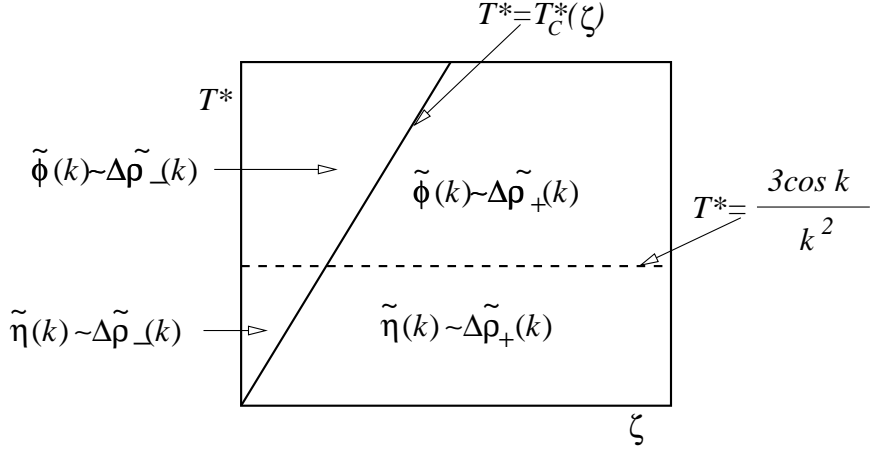}
\caption{Dominant eigenmodes for $\delta, \nu\to 1$ 
($\lambda,Z\to\infty$) in different parts of the phase diagram.  As
discussed in sec.IIIB, the fields $\eta$ and $\phi$ dominate below and
above the dashed line respectively. The relation between the
eigenmodes and $\Delta\rho_{\alpha}$ is given in
Eqs. (\ref{abcinfty3}) and (\ref{abcinfty4}). The neglected
contributions to the OP's are $O(1/Z)$. }
\end{figure}

\section{Summary}
 We performed the MF stability analysis of the disordered phase in the
 PM for the whole range of size and charge asymmetry of the ionic
 species.  At the true instability the {\it average}
 distribution of ions undergos ordering.
As we argue in sec.IID, the instabilities of the disordered
 phase found in MF in fact represent structural lines. The structural
 lines are
 associated with
 ordering of the majority of the {\it instantaneous} states, i.e. with
 pretransitional effects.
  Our results show that periodic
 ordering of the instantaneous states depends mainly on the size
 asymmetry $\delta$. Qualitative
 dependence on the charge asymmetry $\nu$ is found only when the size
 asymmetry is sufficiently large. 

The strongest tendency for periodic ordering is found for
$\delta,\nu\to 1$; in this case the pretransitional effects are
present for a very large temperature range (Fig.19, top). The period
of $\rho_{\alpha}({\bf r}) $, $2\pi/k_b \sim 2\sigma_+$, {\it
decreases} with increasing $\zeta$ (Fig.19, bottom). More detailed
analysis in Ref.\cite{ciach:06:0} shows that for small $\zeta$ the
most probable structure has the form of the bcc crystal formed by the
large ions that are immersed in the cloud of the counterions.
   When both asymmetry parameters decrease the temperature range for
   which the periodic ordering occurs decreases very rapidly
   (Figs.18,14,8). For $\delta,\nu=0.9$, $T^*$ at the maximum of the
   structural line is 500 times lower than for $\delta=0.998,
   \nu=0.9998$ (Figs. 14 and 19). Physical systems corresponding to
   this range of asymmetry parameters include highly charged colloids
   in the presence of one kind of counterions and no coions.

In the other extreme case of $\delta\nu\to -1$ the periodic ordering
is suppressed entirely.  This case is unphysical, and will not be
discussed here.

For $0.9>\delta>0.4$ the periodic ordering depends on the size
asymmetry rather weakly,
 as long as $\nu$ is sufficiently large (Figs. 8 and 14). However,
 when $\nu$ decreases and becomes negative (the charge at the smaller
 ion is not sufficiently small), $T^*$ and $\zeta$ corresponding to
 the ordering both decrease to very small values (Figs. 12 and
 15). Nearly neutral clusters followed by voids are formed in this
 case, and the period of the structure, $2\pi/k_b$, is an {\it
 increasing} function of $\zeta$ (Figs. 11, 13). Moreover, the
 corresponding structural line and the MF gas-liquid spinodal are
 close to each other on the phase diagram (Figs. 12 and 15).  In
 future studies the binodal lines should be determined in MF and
 beyond to verify whether the complex ordered structures with large
 periods correspond to stable crystal phases, or whether they reflect
 a tendency for formation of aggregates in the fluid phase.
 Physical systems corresponding to this range of asymmetry may include
 globular proteins, organic ions,
 and some room-temperature ionic liquids.

For the large and moderate size asymmetry discussed above
($\delta>0.4$) the structural line assumes a pronounced maximum for
$\zeta\approx 0.15$ (Figs. 8, 12, 14, 15, 18). It is remarkable that
the periodic ordering induced by the Coulomb interactions is most
efficient for small volume fractions, whose
 range, $0.1<\zeta< 0.3$, is almost independent of $\delta$ (Figs. 8,
 12, 14, 15, 18).

 When $\delta$ further decreases, the maximum of the structural
 line becomes less and less pronounced, and finally dissappears for
 $\delta\approx 0.4$ ($\lambda<2.3$).
 $T^*$ at the structural line becomes a monotonically increasing
 function of $\zeta$, and the periodic ordering is more efficient at
  large volume fractions (Fig.5). The period of the ordered
 structure is nearly {\it independent} of $\zeta$ (sec.VIA). 
 The same qualitative behavior is found in our MF approximation for the
 RPM~\cite{ciach:00:0}. By analogy with the RPM~\cite{ciach:06:2}
 we expect that beyond MF the ordered phase corresponds to a hard
 ionic crystal. The crystallization line is expected at much lower
 $T^*$ and much higher $\zeta$ compared to the
 structural line~\cite{ciach:06:2}.
Physical systems for this range of asymmetry include molten
 salts and electrolytes.

Finally let us  focus on the separation between two uniform phases.
 We found that some of the effects of the size- and charge asymmetry
 are correctly predicted already on the MF level of our theory
 (sec.V). However, on the MF level the gas-liquid separation is
 preempted by the instability with respect to the periodic ordering,
 except from the case where such ordering is suppressed by geometrical
 constrains. This fact agrees with earlier MF results for the
 RPM~\cite{ciach:00:0,ciach:01:0,ciach:06:1}. In fact, on the MF level
 no phase separation is found in the RPM; the phase separation is
 induced by the charge-charge
 correlations
~\cite{stell:95:0,ciach:00:0,ciach:01:0,ciach:06:1,patsahan:04:0}. Studies
 beyond MF are required to determine for what parts of the parameter
 space $(\zeta,T^*;\delta,\nu)$ the phase separation is indeed
 preempted by the crystallization, and how the critical point varies
 with $\delta$ and $\nu$. Returning to the clustering associated with
 the phase separation we should note that phenomena occurring at the
 very short length-scale cannot be accurately described by the
 coarse-grained theory. It is not clear to what extent the clustering
 of ions is associated with the short-distance behavior of the
 correlation functions, and to what extent it is induced by
 collective phenomena. Studies beyond the stability analysis of
 $\Omega^{MF}$ may shed light on this question.

Short-range interactions of different
origin (including solvet-induced effective interactions between solute
molecules) also play an important role for the phase
behavior~\cite{ciach:05:0}. Such short-range interactions can be
included in the field-theoretic approach by supplementing the energy
contribution to $\Omega^{MF}$ in Eq.(\ref{bolfac}) with additional terms.

\begin{acknowledgments}
GS gratefully acknowledges the support of the Division of Chemical 
Sciences, 
Office of the Basic Energy Sciences, Office of Energy Research, 
US Department of Energy. The work of AC and WTG was partially funded by the 
KBN grant 
 No. 1 P03B 033 26.
\end{acknowledgments}

\section{Appendix A}
The quantities $a_{\alpha\beta}$ in Eq.(\ref{aab}) are obtained from
the explicit expressions for the chemical potentials in the
hard-sphere mixture derived in Ref.
\cite{lebowitz:64:0}. Direct differentiations lead to the 
following expressions
\begin{equation}
\label{a++}
a_{++}=\frac{1}{\rho_{0+}^*} +\frac{\pi r_+^3}{6}\Bigg[
\frac{8}{1-\zeta}+\frac{6r_+^2Y+15r_+X+r_+^3s}{(1-\zeta)^2}+
\frac{18r_+^2X^2+6r_+^3XY}{(1-\zeta)^3}+\frac{9r_+^3X^3}{(1-\zeta)^4}\Bigg]
\end{equation}
with the analogous expression for $a_{--}$,
\begin{equation}
\label{a+-}
a_{+-}=\frac{\pi }{6}\Bigg[\frac{8}{1-\zeta}+
\frac{2r_+r_-(6+r_+r_-)X+8r_+^2r_-^2Y}{(1-\zeta)^2}+
\frac{18r_+^2r_-^2X^2+6r_+^3r_-^3XY}{(1-\zeta)^3}+
\frac{9r_+^3r_-^3X^3}{(1-\zeta)^4}
\end{equation}
where $s$, $\zeta$ and $r_{\pm}$ are defined in Eqs. (\ref{s}), (\ref{zeta}) 
and (\ref{rpm}), and
 following Ref.\cite{lebowitz:64:0} we have introduced the notation 
\begin{equation}
\label{X}
X=\frac{\pi}{6}(r_+^2\rho_{0+}^*+r_-^2\rho_{0-}^*)
=s(1+\delta^2-2\delta\nu)
\end{equation}
\begin{equation}
\label{Y}
Y=\frac{\pi}{6}(r_+\rho_{0+}^*+r_-\rho_{0-}^*)=
s(1-\delta\nu).
\end{equation}

\section{Appendix B}
The functions $b$ and $d$ introduced in Eq. (\ref{1.2}) are defined by

\begin{equation}
\label{bb}
b=(a_{++}Z^{-1}+a_{--}Z)
\cos k\cos (k\delta) +
(a_{++}Z^{-1}-a_{--}Z)
\sin k \sin(k\delta) +
2 a_{+-}\cos k,
\end{equation}
\begin{equation}
\label{d}
d=a_{++}a_{--}-a_{+-}^2=\frac{1}{\rho_{0+}^*\rho_{0-}^*}
\frac{(1+2\zeta)^2}
{(1-\zeta)^4}>0,
\end{equation}

\end{document}